\documentclass[prb, twocolumn, notitlepage]{revtex4-1}
\usepackage{amsmath,amsfonts, amssymb, amsthm, dsfont}
\usepackage{yfonts}
\usepackage{bm}
\usepackage{mathrsfs}
\usepackage{graphicx}
\usepackage{verbatim}
\usepackage{hyperref}
\usepackage{wasysym}

\newcommand{\di}{\mathrm{d}}

\renewcommand{\vec}[1]{{\mathbf #1}}

\renewcommand{\vr}{{\vec{r}}}
\newcommand{\vk}{{\vec{k}}}
\renewcommand{\ol}[1]{\overline{#1}}
\newcommand{\comments}[1]{}
\newcommand{\mb}[1]{\mathbf{#1}}

\newcommand{\coho}[1]{\textswab{#1}}
\newcommand{\cohosub}[1]{\scalebox{0.7}{\textswab{#1}}}

\newcommand{\Ref}[1]{ Ref. [\onlinecite{#1}]}

\def\U{\mathrm{U}(1)}
\def\H{\mathcal{H}}
\def\Z{\mathbb{Z}}

\makeatletter
\def\l@subsection#1#2{}
\makeatother

\begin{document}

\title{Fermionic Lieb-Schultz-Mattis Theorems and Weak Symmetry-Protected Phases}

\author{Meng Cheng}
\affiliation{Department of Physics, Yale University, New Haven, CT 06511-8499, USA}
\date{\today}

\begin{abstract}
	The Lieb-Schultz-Mattis (LSM) theorem and its higher-dimensional generalizations by Oshikawa and Hastings establish that a translation-invariant lattice model of spin-$1/2$'s can not have a non-degenerate ground state preserving both spin and translation symmetries. Recently it was shown that LSM theorems can be interpreted in terms of bulk-boundary correspondence of certain weak symmetry-protected topological (SPT) phases. In this work we discuss LSM-type theorems for two-dimensional fermionic systems, which have no bosonic analogs. They follow from a general classification of weak SPT phases of fermions in three dimensions. We further derive constraints on possible gapped symmetry-enriched topological phases in such systems. In particular, we show that lattice translations must permute anyons, thus leading to ``symmetry-enforced'' non-Abelian dislocations, or ``genons''. We also discuss surface states of other weak SPT phases of fermions.
\end{abstract}
\maketitle


\section{Introduction}
Determining the emergent quantum phase in an interacting quantum many-body system is generally a very difficult problem. The celebrated Lieb-Schultz-Mattis-Hastings-Oshikawa (LSMHO) theorem~\cite{LSM, OshikawaLSM, HastingsLSM} points to an exact relation between the microscopic properties and low-energy physics for lattice spin systems: 
if there is an odd number of spin-$1/2$'s per unit cell, there can not be a short-range entangled (SRE) ground state while preserving both $\mathrm{SO}(3)$ and translation symmetries. A symmetric gapped ground state then must be topologically ordered, e.g. a quantum spin liquid~\cite{SavaryReview, RMPQSL}. The LSMHO theorem has been quite valuable in the study of frustrated magnets. Recently, LSMHO theorem has been generalized significantly, to more complicated space groups~\cite{Parameswaran2013, WatanabePNAS, PoPRL2017} and other internal symmetry groups (e.g. a Kramers doublet per unit cell in the presence of time-reversal symmetry, applicable to spin-orbit-coupled materials), as well as to itinerant fermions with charge conservation~\cite{WatanabePNAS}. More recently, LSMHO theorems have also found extensions to magnetic translation symmetries, which lead to LSM-type constraints for symmetry-protected topological (SPT) phases~\cite{Lu2017, Lu_LSMSPT, Yang_LSMSPT}.
On the other hand, when the ground state is topologically ordered preserving all symmetries, the LSMHO theorem can be further refined to place stringent constraints on the symmetry-enriched topological order~\cite{ZaletelPRL2015, ChengPRX2016, Qi_spinonPSG}. For example, in the ``traditional'' setup with spin $\mathrm{SO}(3)$ and translation symmetries, one can show that the ``background charge'' on a lattice must have spin-$1/2$, i.e. a spinon. These results will be collectively referred to as LSM-type constraints.

Many LSM-type constraints can be unified under the theme of bulk-boundary correspondence~\cite{VishwanathPRX2013, Chen2014} of crystalline SPT phases~\cite{ChengPRX2016, ChoPRB2017, JianPRB2018, HuangPRB2017, ThorngrenPRX2018}. Essentially, given a $d$-dimensional lattice system with an internal symmetry group $G$, if there is a projective representation of $G$ in each unit cell, the translation-invariant system can be viewed as the boundary of a stack of 1D SPT phases (going into a ficticious $(d+1)$-th direction), which forms a weak SPT phase protected by $G\times\mathbb{Z}^d$. The well-known constraints on the boundary state of such a weak SPT phase become the statement of a LSM-type theorem. This immediately suggests the following generalization: we can construct a bulk as a stack of 1D fermionic SPT (FSPT) phases~\cite{Fidkowski2011}. End states of 1D FSPT phases are characterized by ``fermionic projective representations'', namely symmetry transformations not commuting with the local fermion parity. Unlike other LSM-type theorems for itinerant electrons known in literature~\cite{WatanabePNAS}, LSM-type constraints obtained this way have no ``Mott'', or bosonic limits, and do not require charge conservation.

In this work we study fermionic LSM-type constraints, focusing on two-dimensional (2D) lattice systems with an internal particle-hole symmetry and translations. After motivating the theorem from the consideration of bulk-boundary correspondence, we state the general fermionic LSM theorems, and present a simple proof of the theorem in the case of a $\Z_2$ particle-hole symmetry and then analyze its implications for possible symmetric gapped phases. We will also consider constraints on gapped surface states of other weak fermionic SPT phases.

\section{Classification of Weak SPT Phases}
In this section we present the classification of weak FSPT phases in three dimensions (3D), whose internal symmetry group is $\Z_2^f\times G$ with $\Z_2^f$ being the fermion parity symmetry, and the full symmetry group is $\Z^3\times \Z_2^f\times G$. The classification is very similar to the bosonic case~\cite{ChengPRX2016}, with three ``weak invariants'':

\begin{enumerate}
	\item There are ``strong'' SPT phases protected just by $G$ alone.   
	\item We may define ``2D SPT per unit length''.  The generators of the bulk states are obtained by stacking 2D SPT layers along the $\hat{i}$-th direction, where ${i}=x,y,z$. They will be referred to as ``type-I'' weak SPT phases.
\item We define ``1D SPT per unit area''. The generators of the bulk states are obtained by packing 1D SPT perpendicular to the $ij$-plane. They will be sometimes referred to as ``type-II'' weak SPT phases.
\item Lastly, there is also ``zero-dimensional SPT per unit volume'', generated by filling the bulk with 0D SPT states (``charges'').
\end{enumerate}

Since we will consider the physics of the surface, only the type-I and type-II invariants are relevant. The most significant invariant is type-II invariant, i.e. 1D SPT per unit area. Let us review the classification of 1D fermionic SPT phases. It is well-known that 1D topological phases are classified by their end states~\cite{PollmannPRB2012, Chen_PRB2011a, ChenPRB2011b, Fidkowski2011, turner2011, Schuch_PRB2011}. For fermionic systems, we distinguish two cases: in a finite system, if there is a topological ground state degeneracy between states with even and odd parity, then even in the absence of any symmetries the system remains nontrivial, i.e. a class D topological superconductor. Otherwise, we can assume that the (symmetry-protected) degenerate ground states all have even fermion parity~\cite{Fidkowski2011}. We can think of the two cases as having odd/even number of Majorana zero modes on each end.

Let us consider the latter case, where there exists a well-defined Fock space on each end~\cite{Fidkowski2011, BultinckPRB2017, KapustinFSPT1D, TurzilloFSPT1D}. FSPT phases in this case are classified by a pair $(\rho, \omega)$ where $\lambda: G \rightarrow \Z_2=\{0,1\}$ is a group homomorphism and $[\omega]\in \H^2[G, \U]$. $\rho$ determines whether the local symmetry action on the boundary is bosonic ($\rho=0$) or fermionic ($\rho=1$) and $\omega$ determines the projective phases of the local symmetry action, which can be modified by stacking with 1D bosonic SPT phases protected by $G$ symmetry~\footnote{When the total symmetry group is a nontrivial extension of $G$ by $\Z_2^f$, there is a further obstruction-vanishing condition on $\rho$ and $\omega$. See e.g. \Ref{TurzilloFSPT1D} for details. We will only consider total symmetry groups taking the form $\Z_2^f\times G$}.

For a concrete example, we set $G=\Z_2=\{1, \mb{g}\}$. Since $\H^2[\Z_2, \U]=\Z_1$, there is one nontrivial phase with $\rho_\mb{g}=1$. A simple physical realization can be found in a system of two identical Kitaev chains, labeled as $\uparrow$ and $\downarrow$. The $\Z_2$ symmetry is generated by $(-1)^{N_\downarrow}$.

On one end there are two Majorana modes, $\gamma_\uparrow$ and $\gamma_\downarrow$. Under $\mb{g}$ they transform as
\begin{equation}
	\gamma_\uparrow \rightarrow \gamma_\uparrow, \gamma_\downarrow \rightarrow -\gamma_\downarrow.
	\label{}
\end{equation}
Locally this symmetry action can be implemented by a unitary $U_\mb{g}=\gamma_\uparrow$. Obviously the only mass perturbation $i\gamma_\uparrow\gamma_\downarrow$ breaks the $\Z_2$ symmetry. We can also define a complex fermion mode $c=\frac{\gamma_\uparrow-i\gamma_\downarrow}{2}$, then $\Z_2$ acts as a particle-hole (or charge-conjugation) transformation: $c\rightarrow c^\dag$. We will call such a boundary fermionic mode a fermionic projective representation of $G$.

We will also consider 2D SPT per unit length. For $\Z_2^f\times G$ with $G$ being a unitary group, a complete classification of 2D fermionic SPT phases has been established~\cite{GuWen, Cheng_fSPT, BhardwajJHEP2017}. In this case the classification data is a triple $(\rho, \nu, \omega)$ where $\rho: G \rightarrow \Z_2=\{0,1\}$ is again a group homomorphism, $\nu$ is a $2$-cocycle in $\H^2[G, \Z_2]$ and $\omega$ now is a $3$-cochain valued in $\U$. Physically, $\rho$ determines whether a $\mb{g}$-defect is Majorana-type or not (i.e. whether it carries a Majorana zero mode). We will only consider trivial $\rho$ in this work. $\nu$ encodes the projective fusion rules of symmetry defects:   
\begin{equation}
	0_\mb{g}\times 0_\mb{h}=\nu(\mb{g,h})\times 0_\mb{gh}.
	\label{}
\end{equation}
Here $\nu(\mb{g,h})=1,f$. There is also an obstruction-vanishing condition for $\nu$ and $\omega$~\cite{GuWen, Cheng_fSPT}.

We believe the results can also be obtained from a general classification of fermionic SPT phases, such as spin cobordism~\cite{Kapustin2015b}, with an internal symmetry group $\Z^3\times G$. As long as the classification is given by a generalized cohomology theory, a Kunneth-type decomposition exists and produces the classification discussed in this section (see \Ref{Xiong}). We will come back to this point in Sec. \ref{sec:discussion}. 

\section{Fermionic LSM Theorems}
Motivated by the perspective that views LSM-type theorems as a result of bulk-boundary correspondence, we propose fermionic LSM-type theorems: 

\vspace{3mm}
\noindent\fbox{%
    \parbox{\columnwidth}{%
In a $d$-dimensional lattice, if the degrees of freedom in a unit cell transform as a nontrivial fermionic projective representation of an internal symmetry group $G$, and the Hamiltonian preserves both $G$ symmetry and translation symmetries, the ground state can not be gapped and non-degenerate at the same time without breaking symmetries. 
    }%
}
\vspace{3mm}

Let us formally define what is a fermionic projective representation, which is essentially the boundary state of a 1D FSPT phase.
Consider the local Hilbert space of one unit cell. For each $\mb{g}\in G$, let $U_\mb{g}$ denote the unitary representation of $\mb{g}$ on the Hilbert space. Consider the commutator between $U_\mb{g}$ and the local fermion parity $(-1)^{N_f}$:
\begin{equation}
	U_\mb{g}^{-1}(-1)^{N_f}U_\mb{g}= (-1)^{N_f + \rho(\mb{g})}.
	\label{}
\end{equation}
If $\rho(\mb{g})=1$, $U_\mb{g}$ changes the local fermion parity. Clearly $\rho(\mb{g})$ defines a group homomorphism from $G$ to $\Z_2=\{0, 1\}$. The fermionic projective representation is nontrivial if $\rho$ is not trivial.

\subsection{A fermionic LSM theorem with unitary $\Z_2$ symmetry}
The main example that we will study is $G=\mathbb{Z}_2$. In this case, the fermionic projective representation of $G$ is basically a single fermion mode $c$ with a particle-hole symmetry $c\rightarrow c^\dag$.  Thus we are led to consider a lattice model of spinless fermions with exact particle-hole symmetry, and one particle-hole doublet per unit cell. For example, the Hamiltonian may contain purely imaginary hopping terms $\pm ic_\vr^\dag c_{\vr'}$, as well as interactions such as $(n_\vr-1/2)(n_{\vr'}-1/2)$. Notice that the symmetry fixes the average density $\langle n\rangle$ of fermions to be $1/2$, i.e. half filling. So if the fermion number is conserved, the LSMHO theorem already rules out a trivially gapped ground state. However, if the $\U$ symmetry is broken, without additional symmetries it is certainly possible to have a trivial gapped phase even when the average density is fractional. For example, we can simply form a stack of Kitaev chains. 

We now present a proof of the claimed fermionic LSM theorem. Without loss of generality, we consider a square lattice of size $N_x\times N_y$, where fermions obey {periodic} or anti-periodic boundary conditions in both directions:  
\begin{equation}
	c_{x+N_x,y}=B_x c_{xy}, c_{x, y+N_y}=B_yc_{xy}.
	\label{}
\end{equation}
Here $B_{x,y}=\pm 1$.

The $\mathbb{Z}_2$ particle-hole symmetry $\mb{g}$ is generated by the following unitary
\begin{equation}
	U_\mb{g}= \prod_{\vr}(c_\vr+c_\vr^\dag).
	\label{}
\end{equation}
Notice that $U_\mb{g}^2=(-1)^{\frac{N_s(N_s-1)}{2}}$, where $N_s=N_x N_y$ is the number of sites. In the following we define $\gamma_\vr=c_\vr+c_\vr^\dag$.

If both $N_x$ and $N_y$ are odd, $U_\mb{g}$ is fermionic and $\{U_\mb{g}, (-1)^{N_f}\}=0$. This is already an indication that there can not be a fully symmetric  SRE state on the torus, since a SRE state should have a unique ground state on any closed manifold and should not know the parity of $N_{x/y}$ when $N_{x/y}\gg 1$.  

We may also consider the case when there are an even number of sites and $U_\mb{g}$ is a bosonic operator. 
The translation symmetry then acts on the fermions as follows:
\begin{equation}
	\begin{gathered}
	T_x c_{x,y}T_x^{-1}=c_{x+1,y}, x=1,2,\dots, N_x-1, \\
	T_x c_{N_x,y}T_x^{-1}=B_xc_{1,y}.
	\end{gathered}
	\label{}
\end{equation}

Generally we find that
\begin{equation}
	\begin{split}
		T_xU_\mb{g}T_x^{-1}&= T_x\prod_{y=1}^{N_y}\prod_{x=1}^{N_x}\gamma_{x,y}T_x^{-1} \\
	&= \prod_{y=1}^{N_y}B_x\gamma_{2,y}\dots\gamma_{N_x,y}\gamma_{1,y} \\
	&= \big[(-1)^{N_x-1}B_x\big]^{N_y}U_\mb{g}.
	\end{split}
	\label{}
\end{equation}
Similarly we have
\begin{equation}
	T_y U_\mb{g} T_y^{-1} = \big[(-1)^{N_y-1}B_y\big]^{N_x}U_\mb{g}.
	\label{}
\end{equation}
For even $N_x$ and odd $N_y$, we obtain $T_xU_\mb{g}T_x^{-1}U_\mb{g}^{-1}=-B_x$. Therefore, with boundary condition $B_x=1$, there is again at least two-fold ground state degeneracy. This rules out a completely symmetric SRE ground state. The argument presented here is very similar to the one for translation-invariant Majorana models in \Ref{HsiehPRL2016}.

It is also straightforward to show that any quadratic Hamiltonian preserving the symmetries must be gapless. In fact, if we write $c_\vr=\frac{1}{2}(\gamma_\vr-i\eta_\vr)$ where $\gamma_\vr$ and $\eta_\vr$ are Majorana operators, the $U_\mb{g}$ symmetry acts as $\gamma_\vr\rightarrow \gamma_\vr, \eta_\vr \rightarrow -\eta_\vr$. Thus at quadratic level, $\{\gamma_\vr\}$ and $\{\eta_\vr\}$ are decoupled. A translation-invariant Majorana model with one Majorana per site is always gapless since the single-particle dispersion has to be an odd function of the lattice momentum, which means the dispersion must vanish (i.e. gap closing) at the zero momentum.

One can easily construct various symmetry-breaking ground states. For example, on a square lattice we can simply form a charge density wave with ordering vector $(\pi,\pi)$.
Alternatively, we can keep the $\Z_2$ particle-hole symmetry by forming ``bond'' density waves, pairing $\gamma_\vr$ with $\gamma_{\vr'}$, and $\eta_\vr$ with $\eta_{\vr'}$. In the following we present an example of an interacting symmetric gapped phase.

\subsubsection{Coupled-wire construction of a symmetric gapped phase}
\label{sec:coupled-wire}
We construct an example of a symmetric gapped phase on a square lattice, in a highly anisotropic and strongly-interacting limit.  First turn on the following couplings along $x$:
\begin{equation}
	\begin{split}
		H_0&=\frac{t}{2}\sum_{\vr} (i\gamma_\vr\gamma_{\vr+\vec{x}} + i\eta_\vr\eta_{\vr+\vec{x}})\\
&=t\sum_\vr(ic_\vr^\dag c_{\vr+\vec{x}}+\text{h.c.})
	\end{split}
	\label{}
\end{equation}
The single-particle spectrum is $E_{\vk} = 2\sin k_x$. We obtain a stack of gapless chains indexed by $y$ and each of them is a $c=1$ free fermion in $(1+1)$. To describe the low-energy physics, we follow the standard bosonization approach~\cite{giamarchi2003} and linearize the spectrum around Fermi points. Within this approximation, each chain has a right-moving mode $\psi_R$ at $k_x=0$ and left-moving $\psi_L$ at $k_x=\pi$:
\begin{equation}
	H_0=\sum_y\int \di x\, [\psi_{Ry}^\dag (-i\partial_x) \psi_{Ry} + \psi_{Ly}^\dag i\partial_x \psi_{Ly}].
	\label{eqn:h0}
\end{equation}

Under the particle-hole symmetry, the chiral fermion fields transform as
\begin{equation}
	\mb{g}:	\psi_{\lambda y}\rightarrow \psi_{\lambda y}^\dag, \lambda=R,L.
	\label{}
\end{equation}
Under translations they transform as
\begin{equation}
	\begin{split}
	T_x&: \psi_{Ry}\rightarrow \psi_{Ry}, \psi_{Ly}\rightarrow -\psi_{Ly}\\
	T_y&: \psi_{\lambda y}\rightarrow \psi_{\lambda, y+1}.
	\end{split}
	\label{}
\end{equation}
We then bosonize $\psi_{R/L,y}\sim e^{i\phi_{R/L,y}}$.

\begin{figure}[tpb]
	\centering
	\includegraphics[width=\columnwidth]{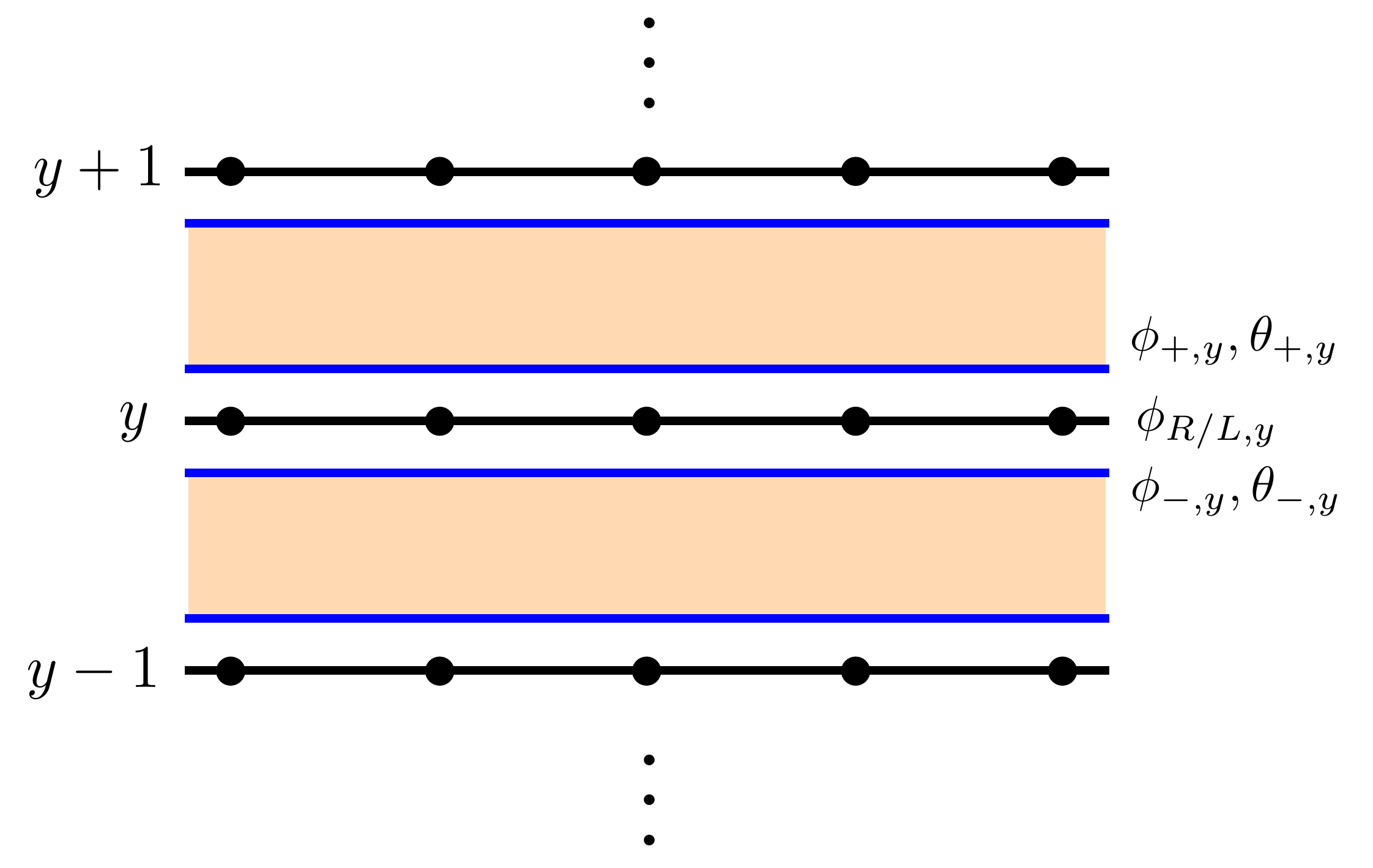}
	\caption{Illustration of the wire construction. The shaded regions are the inserted plates of $\Z_4$ gauge theories.}
	\label{fig:wires}
\end{figure}

To construct a symmetrically gapped phase, the chains must be coupled by interactions. For simplicity of the presentation, we follow an analogous construction in \Ref{MrossPRL2016}, inserting plates of $\mathbb{Z}_4$ gauge theories between neighboring chains (see Fig. \ref{fig:wires}). The bulk of a $\mathbb{Z}_4$ theory can be described by an Abelian Chern-Simons theory with K matrix $K=\begin{pmatrix} 0 & 4\\ 4 & 0\end{pmatrix}$. Anyonic quasiparticles are generated from the $\Z_4$ charge $e$ and flux $m$, as well as their bound states. Correspondingly, edge modes of the $\Z_4$ gauge theory are Luttinger liquid with the following Lagrangian:
\begin{equation}
	\mathcal{L}_\text{edge}=\frac{8}{4\pi}\partial_t\phi\partial_x\theta - \dots
	\label{}
\end{equation}
Here $\phi, \theta$ are compact bosonic fields.
The bulk-edge correspondence identifies $e$ with $e^{i\phi}$, and $m$ with $e^{i\theta}$.

Because of the translation invariance along the $y$ direction, we can focus on the $y$-th chain $\psi_{R/L,y}$ and the two sets of edge mode coming from adjacent plates: $(\phi_\pm, \theta_\pm)$ from the plate between $y$ and $y\pm 1$. The following interactions are turned on~\cite{MrossPRL2016}:
\begin{equation}
	\begin{split}
	H_\text{int}&=-\int\di x\, \Big[U_1\big[(\psi_{R}\psi_L)^2 e^{4i(\phi_{+}+\phi_-)}+\text{h.c.}\big]\\
		&\quad +U_1\big[(\psi_{R}^\dag\psi_L)^2 e^{4i(\theta_{+}-\theta_-)}+\text{h.c.}\big]\\
		&\quad +U_2\cos 4(\phi_+ - \phi_-)\Big].	
	\end{split}
	\label{}
\end{equation}
The gapping terms explicitly preserve $T_x$ and $T_y$ symmetries. To preserve the $\Z_2$ symmetry, we demand that $\mb{g}$ also acts as charge-conjugation symmetry in the $\Z_4$ gauge theory:
\begin{equation}
	\mb{g}: \phi_\pm \rightarrow -\phi_\pm, \theta_\pm \rightarrow -\theta_\pm.
	\label{eqn:Z2action}
\end{equation}
$\mb{g}$ takes $e$ to $\bar{e}$, $m$ to $\bar{m}$ in the $\Z_4$ topological order. Notice that the interactions also preserve $\U$ charge, if $e$ carries a half electric charge. We assume that $U_1, U_2$ are large so the system becomes fully gapped.

Now we have a symmetric, gapped Hamiltonian. We can further check that there is no ground-state degeneracy (except the topological one) from the gapping Hamiltonian at $y$, so there can not be any spontaneous symmetry breaking~\cite{WangPRB2013}. In addition, we have also checked that there is no string operator connecting neighboring wires (which is still a local operator) that transforms under the symmetry and acquires a finite expectation value~\footnote{This is the ``weak symmetry breaking'' phenomenon discussed in \Ref{WangPRB2013}}.
The bulk topological order has been analyzed in \Ref{MrossPRL2016} and is just the $\mathbb{Z}_4$ topological order. However, the symmetry actions on anyons are highly nontrivial, which we determine in the following.

In the gapped phase, the following fields acquire nonzero expectation values:
\begin{equation}
	\begin{gathered}
	O_{1y}=\phi_{Ry}+\phi_{Ly}+2(\phi_{+,y}+\phi_{-,y})\\
	O_{2y}=\phi_{Ry}-\phi_{Ly}-2(\theta_{+,y}-\theta_{-,y})\\
	O_{3y}=\phi_{+,y}-\phi_{-,y}\\
	\end{gathered}
	\label{}
\end{equation}
With our choice of the Hamiltonian, we have $\langle O_{1,2}\rangle = 0,\pi$, and $O_{3j}=0, \pm \pi/2, \pi$. Naively, under $T_x$ we find that $\langle O_1\rangle$ becomes $\langle O_1\rangle+\pi$, and the same to $O_2$. Because $O_1$ and $O_2$ are not local operators, this can be fixed by applying gauge transformations. 

Generally, to be consistent with the topological order,  we must have $\phi_{+,y}, \theta_{+,y}$ and $\phi_{-,y+1}, \theta_{-, y+1}$ transform consistently since they come from edge modes bounding the same bulk. To be precise, we assume that under $T_x$,
\begin{equation}
	\begin{gathered}
	\phi_{\pm,y}\rightarrow \phi_{\pm,y}+\alpha_\pm(y),\\
	\theta_{\pm,y}\rightarrow \theta_{\pm,y}+\beta_\pm(y),
	\end{gathered}
	\label{eqn:gauge-trans}
\end{equation}
where the phases $\alpha_\pm(y)$ and $\beta_\pm(y)$ satisfy
\begin{equation}
	\alpha_+(y) = \alpha_-(y+1), \beta_+(y) = - \beta_-(y+1).
	\label{}
\end{equation}
So that the upper and lower edges of the same plate can be ``glued'' by $\cos 4(\phi_{+,y}-\phi_{-,y+1})$ and $\cos 4(\theta_{+,y}+\theta_{-,y+1})$ without any symmetry breaking.

Demanding that $O_1,O_2$ and $O_3$ are invariant with these gauge transformations, we can easily find $\alpha_\pm(y)=\frac{\pi}{4}$ and $\beta_\pm(y)=\mp\frac{\pi}{4}$.


Let us analyze how $T_y$ acts on anyons.  First, because $\phi_+ = \phi_-$, the $e$ anyons can simply tunnel between different plates. Thus $T_y$ acts trivially on $e$ (aside from moving it along $y$). The situation for $m$ anyons is quite different.  Notice that
\begin{equation}
	O_1-O_2 = 2(\phi_{L}+\phi_{+}+\phi_{-}+\theta_{+}-\theta_{-}).
	\label{}
\end{equation}
Therefore, in terms of the original anyons in each plate, $m_y$ is identified with $m_{y+1}e_{y+1}^2f$, where $f$ represents the physical fermion. This means that in the 2D phase, the actual $m$ anyon has the following identification:
\begin{equation}
	m=
	\begin{cases}
		m_y & y \text{ is even}\\
		m_ye_y^2f & y \text{ is odd}
	\end{cases}
	\label{eqn:m2D}
\end{equation}
Microscopically, $m_y\sim e^{i\theta_{\pm, y}}$ becomes $m_{y+1}\sim e^{i\theta_{\pm,y+1}}$ under the $T_y$ translation. Comparing with Eq. \eqref{eqn:m2D}, we find that $T_y$ acts on anyons as
\begin{equation}
	T_y: e\rightarrow e, m\rightarrow me^2f.
	\label{eqn:Tyaction}
\end{equation}
It is easy to see that the $T_x$ translation does not permute anyons. These results agree with the analysis in \Ref{MrossPRL2016}.

So far we have determined how anyon types are permuted under the symmetries. It is also important to understand symmetry fractionalization, encoded in the various additional phases appearing in the transformations of $\phi$ and $\theta$. Thus we need to check the commutation relations between the symmetries.
First, let us check the commutation relation between $T_x$ and $T_y$. They apparently commute on $e$. For $m$, we will use a heuristic argument here: under $T_y$, $m$ becomes $me^2f$. Since $e$ acquires an $e^{-\frac{i\pi}{4}}$ phase under $T_x$, $me^2f$ acquires an additional $\pm\frac{\pi}{2}$ phase relative to $m$ under $T_x$ ($\pm$ because the transformation of $f$ under $T_x$ has an ambiguity of a $\pi$ phase).
Therefore, when acting on the $m$ anyon, $T_yT_x$ and $T_xT_y$ differ by a $\pm\frac{\pi}{2}$ phase. 
Intuitively we can understand the non-commutativity between $T_x$ and $T_y$ translations as having a background charge $e$ or $\bar{e}$~\footnote{As we will discuss below, these two values are ``gauge-equivalent''.}, such that when moving $m^2$ around a unit cell one gets a $-1$ Berry phase. This is well-defined because $m^2$ is invariant under $T_{x/y}$.  

A similar calculation shows that $T_x$ and $U_\mb{g}$ do not commute when acting on anyons locally. We find that $U_\mb{g}$ and $T_x$ differ by a $\frac{\pi}{2}$ phase on both $e$ and $m$.

\subsection{Fermionic LSM theorems with time-reversal $\Z_2^\mb{T}$ symmetry}
 We can replace the unitary particle-hole symmetry with an anti-unitary one, i.e. time-reversal symmetry $\mb{T}$. For one complex fermion per site, there are two possibilities:
\begin{enumerate}
	\item $\mb{T}^2=\mathds{1}$. In terms of Majorana operators, they transform as
		\begin{equation}
			\mb{T}:\gamma\rightarrow \gamma, \eta\rightarrow \eta.
			\label{}
		\end{equation}
		Or $\mb{T}: c\rightarrow c^\dag$. This kind of transformation forbids all quadratic terms in the Hamiltonian.
	\item $\mb{T}^2=P_f$. In terms of Majorana operators, they transform as
		\begin{equation}
			\mb{T}:\gamma\rightarrow \eta, \eta\rightarrow -\gamma.
			\label{}
		\end{equation}
		This is known as a ``Majorana doublet''.  The complex fermion transforms as $c\rightarrow ic^\dag$. Thus hopping terms like $c_i^\dag c_j$ are not allowed, and only real pairing terms can appear at quadratic level:
		\begin{equation}
			\mb{T}: \Delta c_ic_j\rightarrow \Delta^* c_j^\dag c_i^\dag.
			\label{}
		\end{equation}
\end{enumerate}
In both cases, we conjecture that a SRE ground state preserving all symmetries is forbidden.

\subsection{Majorana LSM-type theorem}
\label{sec:majorana}
We can also consider a LSM-type theorem without any internal symmetries (except the fermion parity conservation), in a translation-invariant lattice with an odd number of Majorana modes per site, e.g. a triangular vortex lattice in a $p_x+ip_y$ superconductor~\footnote{One subtlety is that in such a vortex lattice, the natural Majorana hopping model has magnetic translation symmetry.}. \Ref{HsiehPRL2016} showed that such a lattice model defined on even by odd torus must have degenerate ground states, protected by the anti-commuting algebra of translation and total fermion parity.

\subsection{LSM-type theorems for other space symmetries}
\Ref{PoPRL2017} found the most general LSM-type constraints for 2D magnets, starting from three concrete conditions, known as ``Bieberbach'' no-go, mirror no-go and rotation no-go. 

We conjecture that a general ``Bieberbach'' no-go holds:  if there is a nontrivial fermionic projective representation in the ``fundamental domain'', a symmetric SRE state does not exist. Here a ``fundamental domain''  is a region which tiles the plane under the action of translation and glide symmetries.

However, the generalization of the rotation no-go is more subtle for fermions, as the argument in \Ref{PoPRL2017} is no longer sufficient to exclude SRE ground states preserving all symmetries. A systematic study will be presented elsewhere~\cite{flsm}.

\section{Constraints on 2D Symmetry-Enriched Topological Phases}
In this section we develop fermionic LSM-type constraints for gapped topological phases, generalizing those for bosonic systems obtained in \Ref{ZaletelPRL2015} and \Ref{ChengPRX2016}. We will first develop the necessary formalisms to describe fermionic SET phases and symmetry defects, following the treatment in \Ref{SET}. We should note, however, that our analysis does not cover the most general fermionic SET phases. Indications will be given whenever simplifying assumptions are made.
 The completely general theory will be left for future works.

Throughout this section we assume that the total symmetry group of the system is $\Z_2^f \times G$.

\subsubsection{Algebraic theory of fermionic topological phases}
The mathematical theory of a general two-dimensional topological phase is known as the algebraic theory of anyons (``anyon model''), or unitary braided tensor category (UBTC)~\cite{kitaev2006}. In an anyon model $\mathcal{C}$, a set of labels $a, b, c, \dots$ represent different anyon types, or topological charges. Among them, there is a unique label ``$1$'' denoting the trivial bosonic local excitations.  The most fundamental property of anyon excitations is their fusion rules:
\begin{equation}
	a\times b=\sum_c N_{ab}^c c, 
	\label{}
\end{equation}
where the fusion coefficients $N_{ab}^c>0$ are integers.
Next we can also exchange or braid anyons around another. This information is summarized in the so-called $T$ and $S$ matrices. $T_{ab}=\theta_a \delta_{ab}$ is a diagonal matrix, whose diagonal elements are topological twist factors $\theta_a$, or the exchange phase between $a$. Elements of $S$ matrix are related to mutual braiding statistics between anyons.
For bosonic systems, one can further impose the condition of braiding non-degeneracy, or more precisely that the $S$ matrix is unitary, which makes it into a unitary modular tensor category (UMTC).  Physically, it means that one can always distinguish different topological charges by braiding. There are other finer data for anyon models, notably the $F$ and $R$ transformations, which play important roles in the discussions of symmetry-enriched topological phases, and we refer the readers to \Ref{kitaev2006} for a more comprehensive review.

 Gapped fermionic phases can be modeled as a UBTC, where we include the physical fermion as one of the topological charge types, denoted by $f$ in the rest of the paper. The fermion $f$ satisfies $f^2=1$ and $\theta_f=-1$, but braids trivially with every other anyon. The UBTC is no longer modular, but only up to the physical fermion. We can thus form ``super topological charges'', consisting of a doublet 
\begin{equation}
	\hat{a}=\{a, a\times f\},
	\label{}
\end{equation}
where $a$ is a topological charge in the UBTC~\footnote{One can show that $a\neq a\times f$. Otherwise the ribbon identity if a UBTC implies $R^{a,f}_{a\times f}R^{f,a}_{a\times f}=-1$, contradicting the fact that $f$ is a physical fermion}. The UBTC that describes fermionic systems is ``super-modular''~\cite{FMTC}, in the sense that braiding is still non-degenerate for supercharges. Equivalently, one can factorize the $S$ matrix as
 \begin{equation}
	 S=\hat{S}\times \frac{1}{\sqrt{2}}
	 \begin{pmatrix}
		 1 & 1\\
		 1 & 1
	 \end{pmatrix},
	 \label{}
 \end{equation}
 where $\hat{S}$ is unitary.~\footnote{However, it is worth emphasizing that the factorization of the $S$ matrix does not mean that the UBTC can be factorized into a UMTC and $\{1,f\}$. In general they do not.}

 Every fermionic topological phase modeled by a UBTC $\mathcal{C}_0$ can be embedded into a larger bosonic one $\mathcal{C}$, where $f$ becomes an emergent fermion. Physically, this can be done by ``gauging the fermion parity'', namely coupling the fermionic system to a $\Z_2$ gauge field sourced by the fermions. This gauging is not unique, but we always consider those with the minimal total quantum dimension equal to $\sqrt{2}\mathcal{D}_0$, where $\mathcal{D}_0$ is the quantum dimension of $\mathcal{C}_0$. Such a bosonic topological phase is called the ``modular extension'' of $\mathcal{C}_0$. Even with this condition, there are always $16$ distinct modular extensions of a given $\mathcal{C}_0$, corresponding to stacking 2D topological superconductors before gauging~\cite{kitaev2006, FMTC}. The modular extension can be written as $\mathcal{C}=\mathcal{C}_0\oplus \mathcal{C}_1$, where $\mathcal{C}_1$ consists of fermion parity fluxes which have $-1$ mutual braiding phase with $f$.

\subsubsection{Symmetry action on anyons}
\label{sec:sym-anyons}
We describe the symmetry actions on anyons using the formalism of $G$-crossed braided tensor category~\cite{SET, Tarantino_SET}. For simplicity of the presentation, we only consider unitary $G$ in the following and assume that the fusion multiplicity is always $0$ and $1$.

Let $\mathcal{C}_0$ be the UBTC that describes a fermionic topological phase. Following \Ref{SET}, we define a topological symmetry group Aut$(\mathcal{C}_0)$, consisting of all permutations $\varphi: \mathcal{C}_0\rightarrow \mathcal{C}_0$ under which all physical properties remain invariant, e.g.
\begin{equation}
	\theta_a = \theta_{\varphi(a)}, S_{ab}=S_{\varphi(a), \varphi(b)}.
	\label{}
\end{equation}
Notice that here the definitions of symmetries involve the anyon types directly, not just the supercharge types, i.e. $a$ and $\varphi(a)$ should have \emph{identical} topological twists. For a given permutation $\varphi$, it may act nontrivially on anyon fusion spaces, in order to preserve $F$ and $R$ transformations of the UBTC: 
\begin{equation}
	\varphi(|a,b;c\rangle)=u^{a'b'}_{c'}|a',b';c'\rangle.
	\label{}
\end{equation}
Here we assume that fusion multiplicities are $0$ or $1$ for simplicity, and therefore $u^{a'b'}_{c'}$ are phase factors. In general they are unitary transformations. In particular, the identity permutation can still act on fusion spaces in the following way:
\begin{equation}
	|a,b;c\rangle \rightarrow \frac{\gamma(a)\gamma(b)}{\gamma(c)}|a,b;c\rangle.
	\label{}
\end{equation}
Here $\gamma(a)$ are phase factors. Such  ``trivial'' transformations are called natural isomorphisms.

All allowed permutations form the group Aut$(\mathcal{C}_0)$, which defines the intrinsic symmetry of the emergent topological degrees of freedom. We should however notice that in fermionic systems, there are nontrivial symmetries not captured by Aut$(\mathcal{C}_0)$, which do not permute any anyons in $\mathcal{C}_0$ but permute fermion parity fluxes~\cite{majorana_sym}. We will not consider these symmetries for now.

Given a global internal symmetry group $G$, assuming it is unitary for simplicity, we have a group homomorphism $[\rho]: G\rightarrow \text{Aut}(\mathcal{C}_0)$. Basically, $\rho_\mb{g}$ indicates how a symmetry operation $\mb{g}$ permutes anyons, together with symmetry transformations on fusion spaces. We adopt the following notations from \Ref{SET}:
\begin{equation}
	\begin{gathered}
	\rho_\mb{g}(a)\equiv {}^\mb{g}a,\\
	\rho_\mb{g}|a,b;c\rangle=U_\mb{g}( {}^\mb{g}a, {}^\mb{g}b; {}^\mb{g}c)| {}^\mb{g}a, {}^\mb{g}b;{}^\mb{g}c\rangle.	
	\end{gathered}
	\label{}
\end{equation}
Here $U_\mb{g}( {}^\mb{g}a, {}^\mb{g}b; {}^\mb{g}c)$ represent unitary transformations (and are $\U$ phases in this case). We use $[\rho_\mb{g}]$ to denote the equivalent classes of $\rho_\mb{g}$'s up to natural isomorphisms. The fact that $[\rho]$ forms a group homomorphism means
\begin{equation}
\label{eq:rho_compose}
\kappa_{\bf g,h} \circ \rho_{\bf g} \circ \rho_{\bf h} = \rho_{\bf gh},
\end{equation}
where $\kappa_{\bf g,h}$ is the corresponding natural isomorphism necessary to equate $\rho_{\bf g} \circ \rho_{\bf h}$ with $\rho_{\bf gh}$. More explicitly, we have
\begin{equation}
 \kappa_{\bf g,h} (a,b;c)  =  U_{\bf g}(a,b;c)^{-1}  U_{\bf h}(\,^{\bf \bar g}a,\,^{\bf \bar g}b ;\,^{\bf \bar g}c)^{-1} U_{\bf gh}(a,b;c).
\label{eq:beta_gh}
\end{equation}
Since $\kappa$ is a natural transformation, it takes the following form:
\begin{equation}
	\kappa_{\mb{g,h}}(a,b;c)=\frac{\beta_a(\mb{g,h})\beta_b(\mb{g,h})}{\beta_c(\mb{g,h})}.
	\label{}
\end{equation}

In a gapped phase, let us consider the global symmetry transformation $R_\mb{g}$ corresponding to a group element $\mb{g}$, acting on a physical state containing several anyon excitations $a_1, a_2, \dots, a_n$. Since $G$ is on-site, we expect that the action should be ``localizable'', meaning that the nontrivial action is localized on anyons (on top of the global actions on splitting spaces when anyons are permuted). Formally, we can write
\begin{equation}
	R_\mb{g}|\Psi_{a_1,a_2,\dots, a_n}\rangle = \prod_{j=1}^n U_\mb{g}^{(a_j)} \rho_\mb{g} |\Psi_{a_1,a_2,\dots, a_n}\rangle.
	\label{}
\end{equation}
Here $U_\mb{g}^{(a_j)}$ is a local unitary transformation support on the neighborhood of $a_j$. They form a projective representation of $G$:
\begin{equation}
	U_\mb{g}^{(a)} \rho_\mb{g} U_\mb{h}^{(a)} \rho_\mb{g}^{-1}=\eta_{a}(\mb{g,h}) U_\mb{gh}^{(a)}.
	\label{}
\end{equation}
Associativity of the local unitaries implies that
\begin{equation}
	\eta_a(\mb{g,h})\eta_a(\mb{gh,k})=\eta_{\rho^{-1}_\mb{g}(a)}(\mb{h,k})\eta_a(\mb{g,hk}).
	\label{eqn:sym-loc-1}
\end{equation}

Requiring that $R_\mb{g}R_\mb{h}=R_\mb{gh}$, we find
\begin{equation}
	\prod_{j=1}^n \beta_{a_j}(\mb{g,h})=\prod_{j=1}^n \eta_{a_j}(\mb{g,h}).
	\label{eqn:sym-loc-2}
\end{equation}
In particular, for $N_{ab}^c>0$, we have
\begin{equation}
	\frac{\eta_a(\mb{g,h})\eta_b(\mb{g,h})}{\eta_c(\mb{g,h})}=
	\frac{\beta_a(\mb{g,h})\beta_b(\mb{g,h})}{\beta_c(\mb{g,h})}.
	\label{}
\end{equation}
We define $\eta_a=\beta_a/\omega_a$. The above condition implies
\begin{equation}
	\omega_a(\mb{g,h})\omega_b(\mb{g,h})=\omega_c(\mb{g,h}), N_{ab}^c>0.
	\label{eqn:sym-loc-3}
\end{equation}
One should note that given a homomorphism $[\rho]$ (which fixes $\beta$ up to gauge transformations), Eq. \eqref{eqn:sym-loc-1} and Eq. \eqref{eqn:sym-loc-2} may \emph{not} admit any solutions for $\eta_{a}(\mb{g,h})$. This is captured by a $\H^3$ obstruction class identified in \Ref{SET}. We will proceed assuming that this obstruction vanishes, and there is no further obstruction realizing the corresponding permutation action $\rho$. 
In this case, without loss of generality, we can always choose a representative $\beta_a$ such that
\begin{equation}
	\beta_a(\mb{g,h})\beta_a(\mb{gh,k})=\beta_{\rho^{-1}_\mb{g}(a)}(\mb{h,k})\beta_a(\mb{g,hk}).
	\label{eqn:sym-loc-beta-choice}
\end{equation}
One should remember that this is merely a choice for convenience. With this choice of $\beta_a$, we can easily show that $\omega_a(\mb{g,h})$ satisfies a similar twisted $2$-cocycle condition.

Eq. \eqref{eqn:sym-loc-3} implies that we can always write 
\begin{equation}
	\omega_a(\mb{g,h})=M_{a, \cohosub{w}(\mb{g,h})}^*,
	\label{eqn:M}
\end{equation}
where $\coho{w}(\mb{g,h})$ is an Abelian anyon. This readily follows from the unitarity of $\hat{S}$~\cite{SET}. We denote the set of Abelian anyons by $\mathcal{A}$, which naturally forms an Abelian group with multiplication given by fusion. In the Appendix \ref{app:abelian} we prove that $\mathcal{A}$ can always be written as $\hat{{\mathcal{A}}}\times\{1,f\}$.  Notice that unlike the modular case, $\coho{w}(\mb{g,h})$ is only determined up to the transparent fermion, i.e. only the supercharge $\hat{\coho{w}}(\mb{g,h})$ is fixed by $\omega_a(\mb{g,h})$. 

A straightforward calculation shows
\begin{equation}
	1= M_{a, {}^\mb{g}\cohosub{w}(\mb{h,k})\times\cohosub{w}(\mb{g, hk})\times\ol{\cohosub{w}(\mb{g,h})}\times\ol{\cohosub{w}(\mb{gh,k})}}.
	\label{}
\end{equation}
Therefore we have
\begin{equation}
	{}^\mb{g}\coho{w}(\mb{h,k})\times\coho{w}(\mb{g, hk})\times\ol{\coho{w}(\mb{g,h})}\times\ol{\coho{w}(\mb{gh,k})}\in \{1,f\}.
	\label{}
\end{equation}
In other words, only the supercharge of $\coho{w}(\mb{g,h})$ forms a twisted $2$-cocycle of $G$. 

Let us briefly discuss ambiguities in these quantities. $\eta_a(\mb{g,h})$ is defined up to ``1-coboundaries'', i.e.
\begin{equation}
	{\eta}_a'(\mb{g,h})=\eta_a(\mb{g,h}) \frac{\zeta_a(\mb{gh})}{\zeta_a(\mb{g})\zeta_{\rho_\mb{g}^{-1}(a)}(\mb{h})}
	\label{}
\end{equation}
are physically equivalent to $\eta_a$, by redefining $U_\mb{g}^{(a)}\rightarrow U_\mb{g}^{(a)}\zeta_a(\mb{g})$. Here $\zeta_a(\mb{g})$ are phase factors satisfying $\zeta_a(\mb{g})\zeta_b(\mb{g})=\zeta_c(\mb{g})$ for $N_{ab}^c>0$. This coboundary ambiguity in $\eta_a$ translates into an anyon-valued $1$-coboundary on $\coho{w}$:
\begin{equation}
	{\coho{w}'}(\mb{g,h}) = \coho{w}(\mb{g,h})\times{\coho{z}(\mb{g})\times {}^\mb{g}\coho{z}(\mb{h})}\times\ol{\coho{z}(\mb{gh})}.
	\label{}
\end{equation}
Together, we conclude that equivalence classes of 2-cocycles $[\hat{\coho{w}}]$ valued in Abelian supercharges $\hat{\mathcal{A}}$ are classified by $\H^2_\rho[G, \hat{\mathcal{A}}]$ (the action of $[\rho]$ on $\hat{\mathcal{A}}$ is canonically induced from that of $\rho$ on $\mathcal{A}$).

Each of the $[\hat{\coho{w}}(\mb{g,h})]$ class then represents an equivalence class of projective phases $\eta_a(\mb{g,h})$ characterizing local symmetry actions on anyons. 
Similar to the bosonic case, we refer to $\coho{w}$ as the symmetry fractionalization class. An important remark is in order: when anyons are permuted, $\coho{w}$ should be understood as \emph{torsors}, i.e. starting from a SET phase, we can modify the symmetry fractionalization structure by $\coho{w}$.

\subsubsection{Symmetry defects}
\label{sec:sym-defects}
Symmetry defects are extrinsic objects carrying symmetry fluxes. For each $\mb{g}\in G$, we can introduce $\mb{g}$-defects, going around which a local $\mb{g}$-action is enacted. There are topologically distinct types of $\mb{g}$-defects, 
organized into a $G$-crossed braided category
\begin{equation}
	\mathcal{C}_G^\times = \bigoplus_{\mb{g}\in G} \mathcal{C}_\mb{g},
	\label{}
\end{equation}
where $\mathcal{C}_\mb{g}$ contains all $\mb{g}$-defects $a_\mb{g}, b_\mb{g}, \cdots$. They obey $G$-graded fusion rules:
\begin{equation}
	a_\mb{g}\times b_\mb{h}=\sum_{c_\mb{gh}\in \mathcal{C}_\mb{gh}} N_{a_\mb{g}b_\mb{h}}^{c_\mb{gh}} c_\mb{gh}.
	\label{eqn:g-graded-fusion}
\end{equation}

We will make a simplifying assumption that the symmetry defects do not ``absorb'' physical fermions, i.e. $a_\mb{g}\times f\neq a_\mb{g}$.

\Ref{SET} defines $G$-crossed braiding of defects for bosonic SET phases. While we do not attempt to present the most general theory of $G$-crossed braiding for fermionic SET phases, we will describe in detail an important aspect of $G$-crossed braiding, namely $G$ actions on defects.

Consider a pair of group elements $\mb{g}$ and $\mb{h}$. Suppose in a particular SET phase, $\mb{h}$-defects transform under the $\mb{g}$ symmetry as
\begin{equation}
	{}^\mb{g}(a_\mb{h})= {\rho}_\mb{g}(a_\mb{h}).
	\label{eqn:g-action-defects}
\end{equation}
Here ${\rho}_\mb{g}(a_\mb{h})$ is a defect in the $\mb{ghg}^{-1}$ sector. Following the convention in \Ref{SET}, a counter-clockwise exchange (more precisely, a $R$ transformation) of $a_\mb{g}$ and $b_\mb{h}$ results in $ {}^\mb{g}(b_\mb{h})$ and $a_\mb{g}$, since $b_\mb{h}$ passes through the branch cut of $a_\mb{g}$. Since the exchange can be implemented locally, the total topological charge before and after the $R$ transformation must remain the same.

In the following we consider what happens when we modify the SET structure by a symmetry fractionalization class $[\coho{w}]$. Due to the torsor nature of $[\coho{w}]$, we will assume a ``reference'' SET with $G$-graded fusion rules  in Eq. \eqref{eqn:g-graded-fusion} and $G$ actions given by $\rho$ (see Eq. \eqref{eqn:g-action-defects}), respectively. In the new SET, the fusion rules of defects become
\begin{equation}
	a_\mb{g}\times b_\mb{h}=\coho{w}(\mb{g,h})\times\sum_{c_\mb{gh}\in \mathcal{C}_\mb{gh}} N_{a_\mb{g}b_\mb{h}}^{c_\mb{gh}} c_\mb{gh}
	\label{eqn:defect-fusion}
\end{equation}
to account for the additional projective phases $\omega_a(\mb{g,h})$ when braiding defects around anyons.
The key point is that although the projective phases are determined by $\hat{\coho{w}}$, what appear in the defect fusion rules are $\coho{w}$. 

Let us determine how the symmetry action on defects is modified. Recall that the symmetry action can be implemented by a braid (more precisely, a $R$ transformation). We expect that ${}^\mb{g}(a_\mb{h})$ should differ from ${\rho}_\mb{g}(a_\mb{h})$ by an Abelian anyon:
\begin{equation}
	{}^\mb{g}(a_\mb{h})= {\rho}_\mb{g}(a_\mb{h})\times \coho{b}(\mb{g,h}).
	\label{}
\end{equation}
Now compare the defect fusion rules before and after the braid:
\begin{equation}
	\begin{split}
		a_\mb{g}\times b_\mb{h}&=\coho{w}(\mb{g,h})\times\sum_{c_\mb{gh}\in \mathcal{C}_\mb{gh}} N_{a_\mb{g}b_\mb{h}}^{c_\mb{gh}} c_\mb{gh},\\
		{}^\mb{g}b_\mb{h}\times a_\mb{g}&=\coho{b}(\mb{g,h})\times\coho{w}(\mb{ghg^{-1},g})\times\sum_{c_\mb{gh}\in \mathcal{C}_\mb{gh}} N_{a_\mb{g}b_\mb{h}}^{c_\mb{gh}} c_\mb{gh}.\\
	\end{split}
	\label{}
\end{equation}
In order for them to be equal we find that the ``commutator'' is given by 
\begin{equation}
	\coho{b}(\mb{g,h})=\coho{w}(\mb{g,h})\times\overline{\coho{w}(\mb{ghg^{-1},g})}.
	\label{}
\end{equation}

Symmetry transformations of defects are subject to the following ambiguities: first, for an Abelian anyon $x$ we have
\begin{equation}
	a_\mb{h}\times (x\times \ol{ {}^\mb{h}x})=a_\mb{h}.
	\label{}
\end{equation}
Second, we are free to ``relabel'' charges in a given defect sector, by $a_\mb{h}'=a_\mb{h}\times \varepsilon(\mb{h})$ where $\varepsilon(\mb{h})$ is an Abelian anyon (if $\varepsilon(\mb{h})$ is of the form $x_\mb{0}\times \ol{ {}^\mb{h}x_\mb{0}}$ then the relabeling does not do anything). The symmetry transformation becomes 
\begin{equation}
	{}^\mb{g} (a_\mb{h}')={}^\mb{g}(a_\mb{h}) \times {}^\mb{g}\varepsilon(\mb{h})=\rho_\mb{g}(a_\mb{h}')\times \ol{\varepsilon( {}^\mb{g}\mb{h})}\times {}^\mb{g}\varepsilon(\mb{h}).
	\label{}
\end{equation}
Here one naturally defines $\rho_\mb{h}(a_\mb{g}')=\rho_\mb{h}(a_\mb{g})\times \varepsilon({}^\mb{g}\mb{h})$. We note that these ambiguities are exactly the coboundary degrees of freedom for $\coho{b}(\mb{g,h})$.

\subsubsection{Incorporating translation symmetries}
 Although translations are not internal symmetries, they do preserve locality (i.e. map local operators to local operators), as well as orientation. It is therefore straightforward to formally include translation symmetries into the discussions in Sec. \ref{sec:sym-anyons}. Similarly, we can also discuss ``defects'' of translation symmetries, which are lattice dislocations. The mathematical formulation remains basically identical. We will address a subtlety in the physical interpretation of the action of lattice translation on internal symmetry defects below.

\subsection{Derivation of the LSM-type constraint}
We now derive a generalized LSM-type constraint in a possible gapped symmetric phase. We follow the argument in \Ref{ChengPRX2016}, where it was shown that in a bosonic system, where each unit cell transforms as a projective representation under an internal unitary symmetry $G$, a gapped symmetric topological phase must satisfy
\begin{equation}
	o_{xy}(\mb{g,h})=M_{\cohosub{b}(T_y, T_x), \cohosub{w}(\mb{g,h})} M_{\cohosub{b}(T_x, \mb{g}), \overline{\cohosub{b}(T_y, \mb{h})}},
	\label{}
\end{equation}
assuming no anyons are permuted. Here $[o_{xy}]\in \H^2[G, \U]$ is the factor set that defines the projective representation per unit cell. The argument proceeds by considering moving a $G$ defect around a unit cell. We can identify two contributions in the expression: the first term $M_{\cohosub{b}(T_y, T_x), \cohosub{w}(\mb{g,h})}$ from the background anyon charge $\coho{b}(T_y, T_x)$ which transforms as a projective representation of $G$, and the second from the ``anyonic spin-orbit coupling'' (referring to nontrivial commutation relation between $T_{x/y}$ and $\mb{g}$ when acting on anyons). The physical origin of the second term is that the topological charge of a $\mb{g}$-defect may change under translations. Creation of these additional anyons results in the second phase factor.

In our case, the physical constraint is that the fermion parity in a unit cell changes under the action of the particle-hole symmetry $\mb{g}$. Therefore, it is clear that anyons have to be permuted by some of the symmetries ($\mb{g}$, $T_x$ or $T_y$) to match the additional $f$ fermion that appears under the local symmetry action. Following \Ref{ChengPRX2016}, we introduce a $\mb{g}$-defect into the system, move it around a unit cell, and then apply a local $\mb{g}$ symmetry action to the unit cell in order to restore the original Hamiltonian (removing the branch loop). As the $\mb{g}$-defect is moved, it may change type and leave behind on its path additional anyon charges. We denote the total (Abelian) charge appearing in this process by $\phi_\mb{g}$. The LSM-type constraint essentially says $\phi_\mb{g}=f$.

Let us consider adiabatically transporting a $\mb{g}$-defect where $\mb{g}$ is an internal symmetry. Suppose a unit translation along the $i$-th direction, denoted by $T_i$, acts on defects as
\begin{equation}
	{}^{T_i}(a_\mb{g})=\rho_{T_i}(a_\mb{g}).
	\label{}
\end{equation}
If $\rho_{T_i}$ is nontrivial, na\"ively it might seem that such a translational symmetry action would imply that ${\bf g}$-defects cannot be adiabatically transported in the $\hat{i}$-direction, since the topological charge value carried by the defect must change when it moves. This subtlety was already addressed in \Ref{ChengPRX2016}. For a ${\bf g}$-defect carrying the energetically favored topological charge of $a_{\bf g}$, adiabatically transporting the defect by one unit length in the $\hat{i}$-direction involves extending the defect branch line by one unit length and ending with a Hamiltonian that energetically favors topological charge $\rho_{T_i}(a_\mb{g})$ at the new endpoint of the branch line. For example, if $\rho_{T_i}(a_\mb{g})=a_\mb{g}\times x_0$ where $x_0$ is an Abelian anyon, adiabatically transporting a $a_\mb{g}$ defect by a unit length in the $\hat{i}$-direction involves creating a $x_0-\overline{x_0}$ pair, leaving $x_0$ on the new segment of defect branch line, and fusing $\overline{x_0}$ with the defect to change its topological charge value.

Now we consider the local ${\bf g}$-action on a unit cell by pair creating a ${\bf g}$-${\bf \bar{g}}$ defect pair, adiabatically transporting the ${\bf g}$-defect around a path enclosing one unit cell in a counterclockwise fashion, and then pair annihilating the defects.  As we have mentioned, this process changes the topological charge in the unit cell enclosed by $\phi_\mb{g}$ (which is an anyon, not to be confused with a $\mb{g}$-defect). Because of the torsoring structure in the SET classification, we will actually calculate the \emph{additional} Abelian charge $\Delta\phi_\mb{g}$ accumulated in the unit cell if we modify the SET structure by a fractionalization class $[\coho{w}]$. 

\begin{figure}[t]
	\centering
	\includegraphics[width=\columnwidth]{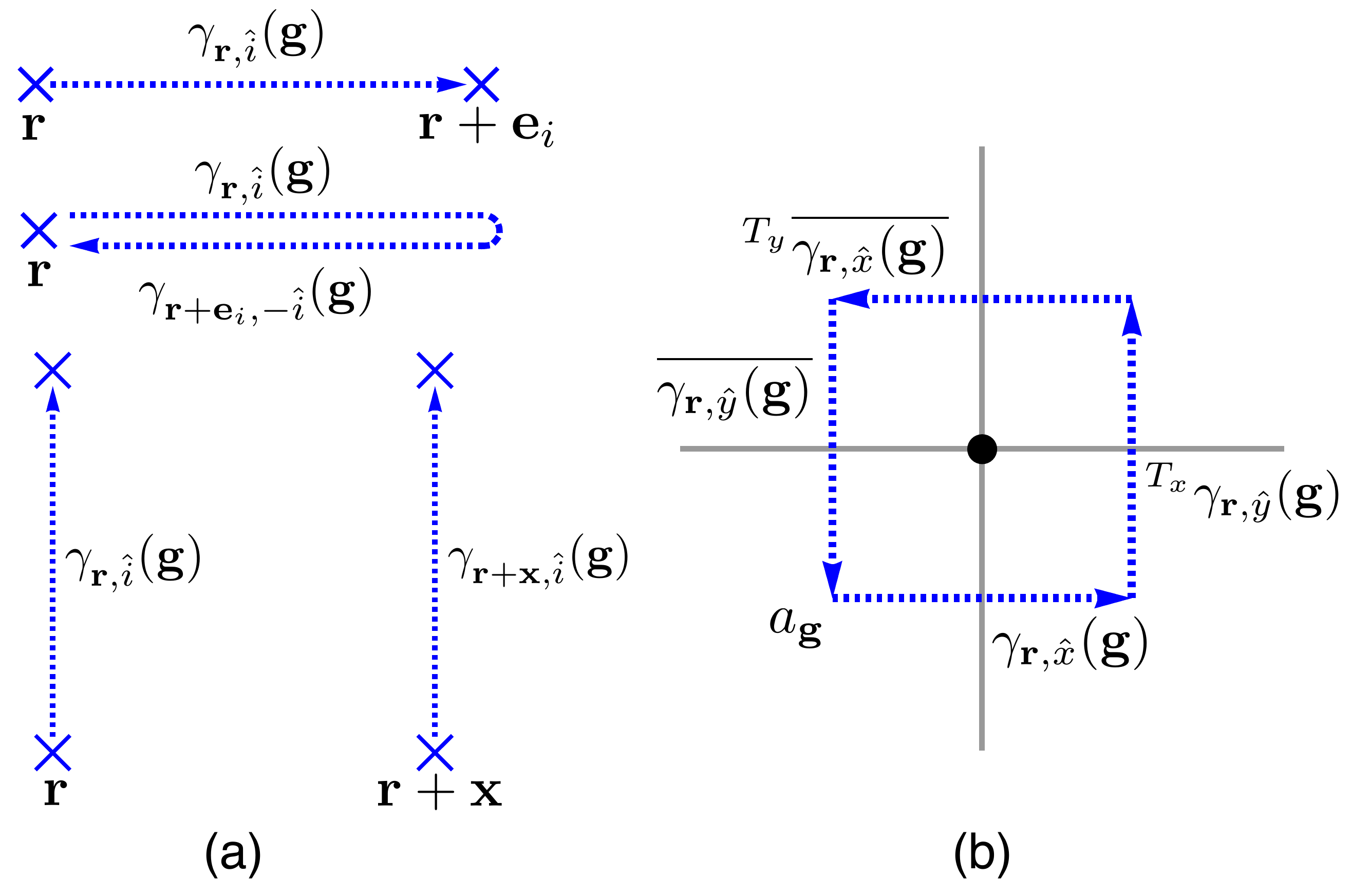}
	\caption{Dashed lines represent defect branch lines. (a) Illustrations of basic relations for $\gamma_{\vr,\hat{i}}(\mb{g})$. (b) Topological charge configurations on defect branch lines after adiabatically transporting a $\mb{g}$-defect around a unit cell.}
	\label{fig:asoc}
\end{figure}

One subtlety which is only present with nontrivial anyon permutation, is that the charge being created when moving a $\mb{g}$-defect is position-dependent. We therefore use $\gamma_{\vr,\hat{i}}(\mb{g})$ to denote the charge created by moving a $\mb{g}$-defect at position $\vr$ by a unit step along the $\hat{i}$-direction, see Fig. \ref{fig:asoc} for illustration. To be precise, again due to the torsoring structure, what we actually calculate is the \emph{additional} Abelian charge associated to the fractionalization class $[\coho{w}]$. We will not repeat this point further.

Before we discuss the actual calculation, let us explain two basic rules:
\begin{itemize}
	\item If we move a $\mb{g}$-defect by a unit length in the $\hat{i}$-direction, and then move it back by a unit length in the $-\hat{i}$ direction so that it returns to the original position, there should be no anyons left. In other words, without loss of generality, we can set
		\begin{equation}
			\gamma_{\vr+\mb{e}_i,-\hat{i}}(\mb{g})\sim\overline{\gamma_{\vr,\hat{i}}(\mb{g})}.
			\label{}
		\end{equation}
	\item Consider moving a $\mb{g}$-defect at position $\vr$ by a unit length in the $\hat{i}$-direction, and the same process but with the $\mb{g}$-defect initially in $\vr+\mb{x}$. These two processes are related by a lattice translation along $\mb{x}$, so we should have 
		\begin{equation}
			\gamma_{\vr+\mb{x},\hat{i}}(\mb{g})\sim{}^{T_\mb{x}}{\gamma_{\vr,\hat{i}}(\mb{g})}.
			\label{}
		\end{equation}
\end{itemize}
Notice that here we write $\sim$ instead of an equality, because one can always absorbs/emits an Abelian anyon of the form ${}^\mb{g} x \times \ol{x}$ where $x\in\mathcal{A}$ to/from a $\mb{g}$-defect. These rules are illustrated in Fig. \ref{fig:asoc}(a).

To be more concrete (without any loss of generality), let us choose the path to be $\mathbf{r}\rightarrow \mathbf{r}+\hat{x}\rightarrow \mathbf{r}+\hat{x}+\hat{y}\rightarrow \mathbf{r}+\hat{y}\rightarrow \mathbf{r}$, as shown in Fig. \ref{fig:asoc}(b).
Keeping track of the topological charge creation and annihilation due to the adiabatic transportation of the ${\bf g}$-defect creating a loop of defect branch line, we have the following: 
\begin{enumerate}
	\item $\mathbf{r}\rightarrow \mathbf{r}+\hat{x}$: the topological charge created on the bottom segment of defect branch line is $\gamma_{\vr,\hat{x}}(\mb{g})$.
	\item $\mathbf{r}+\hat{x}\rightarrow \mathbf{r}+\hat{x}+\hat{y}$: the topological charge created on the right segment of defect branch line is $\gamma_{\vr+\hat{x},\hat{y}}(\mb{g})\sim {}^{T_x}\gamma_{\vr,\hat{y}}(\mb{g})$.
	\item $\vr+\hat{x}+\hat{y}\rightarrow \vr+\hat{y}$: the topological charge created on the upper segment of defect branch line is
		\[
			\gamma_{\vr+\hat{x}+\hat{y}, -\hat{x}}(\mb{g})\sim {}^{T_y}\gamma_{\vr+\hat{x},-\hat{x}}(\mb{g})\sim {}^{T_y}\ol{\gamma_{\vr,\hat{x}}(\mb{g})}.
\]
	\item $\vr+\hat{y}\rightarrow \vr$: the topological charge created on the left segment of the defect branch line is 
		\[
			\gamma_{\vr+\hat{y},-\hat{y}}(\mb{g})\sim\ol{\gamma_{\vr,\hat{y}}(\mb{g})}.
	    \]
\end{enumerate}
The corresponding configuration of topological charges for this ${\bf g}$-defect branch loop is illustrated in Fig.~\ref{fig:asoc}(b). 

All together, an Abelian anyon
\begin{equation}
	\gamma_{\vr,\hat{x}}(\mb{g}) \times {}^{T_x}\gamma_{\vr,\hat{y}}(\mb{g})\times {}^{T_y}\ol{\gamma_{\vr,\hat{x}}(\mb{g})} \times\ol{\gamma_{\vr,\hat{y}}(\mb{g})}
	\label{eqn:asoc-change}
\end{equation}
has been accumulated in the unit cell. We can choose $\vr$ such that
\begin{equation}
	\gamma_{\vr,\hat{i}}(\mb{g})=\coho{b}(T_i,\mb{g}).
	\label{}
\end{equation}

In addition, the $\mb{g}$ action also changes the background charge value:
\begin{equation}
	{}^\mb{g}\coho{b}(T_x, T_y) \times \ol{\coho{b}(T_x, T_y)}.
	\label{eqn:bg-change}
\end{equation}
Note that this result depend on the actual location of the unit cell, since the ``background charge'' can get permuted as well under translations.  However, Eq. \eqref{eqn:bg-change} can be absorbed by a $\mb{g}$-defect so we may ignore this contribution.   

To summarize, we have found that the total change of topological charge in the unit cell is modified by
\begin{equation}
	\Delta\phi_\mb{g}\sim
	{\coho{b}(T_x, \mb{g})}\times {}^{T_y}\overline{\coho{b}(T_x, \mb{g})} \times \overline{\coho{b}(T_y, \mb{g})} \times {}^{T_x}\coho{b}(T_y, \mb{g}).
	\label{eqn:lsm-1}
\end{equation}
We now argue that LSM constraint implies $\Delta\phi_\mb{g}=f$. Let us start from a certain ``reference'' SET, with the same anyon permutation $\rho:G\rightarrow \text{Aut}(\mathcal{C})$. We assume that this reference SET can be realized in a lattice model without a fermionic projective representation per unit cell, so after taking a $\mb{g}$-defect around a unit cell we accumulate no charge: $\phi_\mb{g}=1$. Now we modify the fractionalization class by $[\coho{w}]$ and demand that the resulting SET can be realized in a system with the fermionic LSM constraint.  This is achieved by requiring $\Delta \phi_\mb{g}=f$.

We can in fact make the expression  Eq. \eqref{eqn:lsm-1} more precise. Physically we expect that $\Delta\phi_\mb{g}=1$ or $f$ is ``gauge-invariant'' under coboundary transformations of fractionalization classes. However, the right-hand side of the equation, as given, is not. But recall we are allowed to have additional anyons of the form $x\times {}^{\mb{g}}\ol{x}$. Demanding gauge invariance, the right-hand side can be uniquely fixed: 
\begin{multline}
	\Delta\phi_\mb{g}={}^\mb{g}\coho{b}(T_x, T_y) \times \ol{\coho{b}(T_x, T_y)}\times\\
	{\coho{b}(T_x, \mb{g})}\times {}^{T_y}\overline{\coho{b}(T_x, \mb{g})} \times \overline{\coho{b}(T_y, \mb{g})} \times {}^{T_x}\coho{b}(T_y, \mb{g}).
	\label{eqn:dphi}
\end{multline}

	An interesting corollary of the LSM-type constraint is that $T_x$ or $T_y$ must permute anyons.  If we assume that neither translation permutes anyons, we have that
\begin{equation}
	\Delta\phi_\mb{g}={}^\mb{g}\coho{b}(T_x, T_y) \times \ol{\coho{b}(T_x, T_y)}.
	\label{}
\end{equation}
On the other hand the fermionic LSM constraint requires $\Delta\phi_\mb{g}=f$. So we must find an Abelian background charge $\coho{b}(T_x, T_y)$ such that ${}^\mb{g}\coho{b}(T_x,T_y)=f\times \coho{b}(T_x,T_y)$, which is clearly impossible because for any anyon $a$, ${}^\mb{g}a$ must have the same topological twist $\theta$ as $a$, but $\theta_{f\times a}=-\theta_a$.  Because of the permutation, a lattice dislocation must carry non-Abelian zero modes~\cite{bombin2010, you2012, teo2013}, i.e. they are ``genons''~\cite{barkeshli2012a, barkeshli2013genon}.

Let us check that the D$(\Z_4)$ example constructed in Sec. \ref{sec:coupled-wire} does satisfy the constraint. First we need to calculate the fractionalization data $\coho{b}$. Let us consider $\coho{b}(T_x, T_y)$ as an example. According to the analysis in Sec. \ref{sec:coupled-wire}, $T_xT_y$ and $T_yT_x$ differ by a $\pm i$ phase when acting on $m$. It is also straightforward to check that when they are identical when acting on $e$. In terms of the algebraic data we defined in Sec. \ref{sec:sym-anyons}, this means that
\begin{equation}
	\frac{\omega_e(T_x, T_y)}{\omega_e(T_y, T_x)}=1, \frac{\omega_m(T_x, T_y)}{\omega_m(T_y, T_x)}=\pm i.
	\label{}
\end{equation}
Using the definition in Eq. \eqref{eqn:M} we conclude that $\coho{b}(T_x, T_y)=e$ or $\ol{e}$, which differ by a gauge transformation. Similarly, we have $\coho{b}(\mb{g}, T_x)=em$. Using the transformation rules given in Eq. \eqref{eqn:Z2action} and \eqref{eqn:Tyaction}, Eq. \eqref{eqn:dphi} gives $\Delta\phi_\mb{g}=f$ as expected.

Let us also consider a $T_y$ dislocation $\sigma$. Consider transporting a $m$ anyon around $\sigma$, after which it becomes $me^2f$. In order to conserve topological charge, $e^2f$ must be absorbed by $\sigma$. In other words, $\sigma$ should satisfy $\sigma\times e^2f = \sigma$. Symmetry of fusion coefficients then implies the following fusion rule:
	\begin{equation}
		\sigma_{T_y}\times \ol{\sigma}_{T_y^{-1}} = 1 + e^2 f,
		\label{}
	\end{equation}
	i.e. they are Ising-like defects.

\subsection{Surface States of Type-I Weak Fermionic SPT Phases}
In this section we study gapped surface topological phases of type-I weak SPT phases.

The coupled wire construction presented in Sec. \ref{sec:coupled-wire} can also be used as a model for the surface of a type-I weak fermionic SPT phase, with an internal $\Z_2$ symmetry. The bulk of this SPT is simply a stack of two-dimensional $\Z_2$ fermionic SPT layers. Each layer consists of two Chern insulators, with $C=1$ and $-1$ respectively, and the $\Z_2$ symmetry is the fermion parity of one of the Chern insulators. It is easy to see that the low-energy surface theory is just Eq. \eqref{eqn:h0}, and the $\Z_2$ symmetry acts as
\begin{equation}
	\mb{g}: \psi_{Ry}\rightarrow \psi_{Ry}, \psi_{Ly}\rightarrow -\psi_{Ly},
	\label{}
\end{equation}
i.e. identical to the $T_x$ action at low energy in Sec. \ref{sec:coupled-wire} and  should not be confused with the particle-hole symmetry there.

We use exactly the same construction to obtain a gapped surface. The topological order is still $\mathcal{C}_0=\mathrm{D}(\Z_4)\times\{1,f\}$. The symmetry transformations have been completely worked out in Sec. \ref{sec:coupled-wire}, as long as we interpret $T_x$ as the $\Z_2$ symmetry. The only difference is that a new fractionalization class needs to be considered, namely whether a $\Z_2$-invariant anyon carries a fractional $\Z_2$ charge or not. From Eq. \eqref{eqn:gauge-trans}, we find that indeed the $e$ anyon carries a quarter of $\Z_2$ charge (the cohomology classification is $\mathcal{H}^2[\Z_2, \Z_4]=\Z_2$). Notice that this fractionalization class is absent for the original $T_x$ symmetry since $\mathcal{H}^2[\Z, \Z_4]=\Z_1$.

Let us understand why this SET is anomalous. First we use a flux-fusion anomaly test introduced in \Ref{HermelePRX2016}. Consider inserting a $\Z_2$ flux $0_\mb{g}$ on the surface. Because the $\Z_2$ symmetry does not permute anyons, all $a_\mb{g}$ are Abelian. Thus the defect sector 
\begin{equation}
	\mathcal{C}_\mb{g}=\{a_\mb{g} \equiv 0_\mb{g}\times a\,|\,a\in \mathcal{C}_\mb{0}\}.
	\label{}
\end{equation}

First we argue that the following defect fusion rule must hold:
\begin{equation}
	0_\mb{g}^2=e^{2p+1}m^{2q+1} f^r,
	\label{eqn:defect-fusion-3}
\end{equation}
where $p,q, r$ are integers. This follows from the defect fusion rule Eq. \eqref{eqn:defect-fusion}, and here we give a more heuristic argument:
consider fusing two $0_\mb{g}$ defects. Since $\mb{g}^2=1$, after fusion there is no symmetry defect anymore and therefore $0_\mb{g}^2$ must be an anyon. To determine the anyon type, we braid other anyons around $0_\mb{g}^2$. The braiding phase of an anyon around a $\mb{g}$-symmetry defect is given by the symmetry transformation of the anyon under $\mb{g}$ action. From Eq. \eqref{eqn:gauge-trans}, we know that $m$ picks up a phase $\pm\frac{\pi}{4}$ when transported around a $\mb{g}$-defect, and same with $e$. Therefore, $e$ ($m$) has a braiding phase $\pm\frac{\pi}{2}$ with $0_\mb{g}^2$, which leads to the fusion rule in Eq. \eqref{eqn:defect-fusion-3}.
By relabeling defects $0_\mb{g}\rightarrow 0_\mb{g} m^p e^q$, we can set $p=q=0$ without loss of any generality.

Let us determine how $T_y$ acts in the defect sector. Because $T_y$ commutes with $\mb{g}$, $\mathcal{C}_\mb{g}$ stays the same under $T_y$. Thus we can write
\begin{equation}
	{}^{T_y}(0_\mb{g})=a_\mb{g}
	\label{}
\end{equation}
for some $a\in \mathcal{C}_\mb{0}$.
Therefore
\begin{equation}
	{}^{T_y}(0_\mb{g}^2)=({}^{T_y}0_\mb{g})^2=0_\mb{g}^2 \times a^2=em  a^2f^r
	\label{}
\end{equation}
However, we also know that
\begin{equation}
	{}^{T_y}(0_\mb{g}^2)={}^{T_y}(m f^r)=me^3 f^{r+1}.
	\label{}
\end{equation}
Comparing the two expressions, $a$ must satisfy
\begin{equation}
	a^2=e^2 f.
	\label{}
\end{equation}
This is clearly impossible. In other words, there is no way to take a ``square root'' of the permutation action on $m$.

It is also instructive to consider gauging the fermion parity,  i.e. inserting fermion parity fluxes (i.e. $\pi$ fluxes) into the SET.  Since the surface topological order can be thought as a MTC D$(\Z_4)$ together with trivial fermions, gauging the fermion parity can be done easily: we simply gauge the trivial fermions, and the result is one of Kitaev's 16-fold ways~\cite{kitaev2006}. Without loss of generality, we choose the simplest one of them, namely a $\Z_2$ toric code. We denote the fermion parity flux by $\sigma$. To summarize, the modular extension can be written as
\begin{equation}
	\mathrm{D}(\Z_4)\times\{1, f, \sigma, f\sigma\}.
	\label{}
\end{equation}
Here $\sigma$ is the fermion parity flux and $\theta_\sigma=1$.

We now study how $\mb{g}$ and $T_y$ act in the gauged SET. First let us consider the action $T_y$. To preserve braiding statistics, we must require
\begin{equation}
	M_{ {}^{T_y}\sigma, me^2f}=M_{\sigma, m}=1.
	\label{}
\end{equation}
Because by definition $M_{\sigma, f}=-1$, the left-hand side becomes
\begin{equation}
	M_{ {}^{T_y}\sigma, me^2f}=-M_{ {}^{T_y}\sigma, me^2}.
	\label{}
\end{equation}
Thus $M_{ {}^{T_y}\sigma, me^2}=-1$. Similarly $M_{ {}^{T_y}\sigma, e}=M_{\sigma,e}=1$. These two conditions fix the transformation of $\sigma$: 
\begin{equation}
	{}^{T_y}\sigma=\sigma \times e^2 f^s.
	\label{}
\end{equation}

We turn to the action of $\mb{g}$. $\mb{g}$ does not permute any anyons, and it is easy to see that this is still true after gauging the fermion parity (the only allowed nontrivial permutation for $\mb{g}$ is that $\mb{g}$ takes $\sigma$ to $f\sigma$. This can be thought of as attaching a 2D $\Z_2$ fermionic SPT phase \emph{on the surface}, so can always be undone to produce a surface SET where $\mb{g}$ does not permute anything after gauging). However, we immediately see a problem: we can not consistently assign a $\Z_2$ quantum number to $\sigma$. The reason is that $e^2 f^s$ carries a half $\Z_2$ charge coming from the fractionalization on $e$. Therefore ${}^{T_y}\sigma$ and $\sigma$ necessarily have \emph{distinct} fractional quantum numbers, which is an indication of the anomaly.

\begin{figure}[t]
	\centering
	\includegraphics[width=0.8\columnwidth]{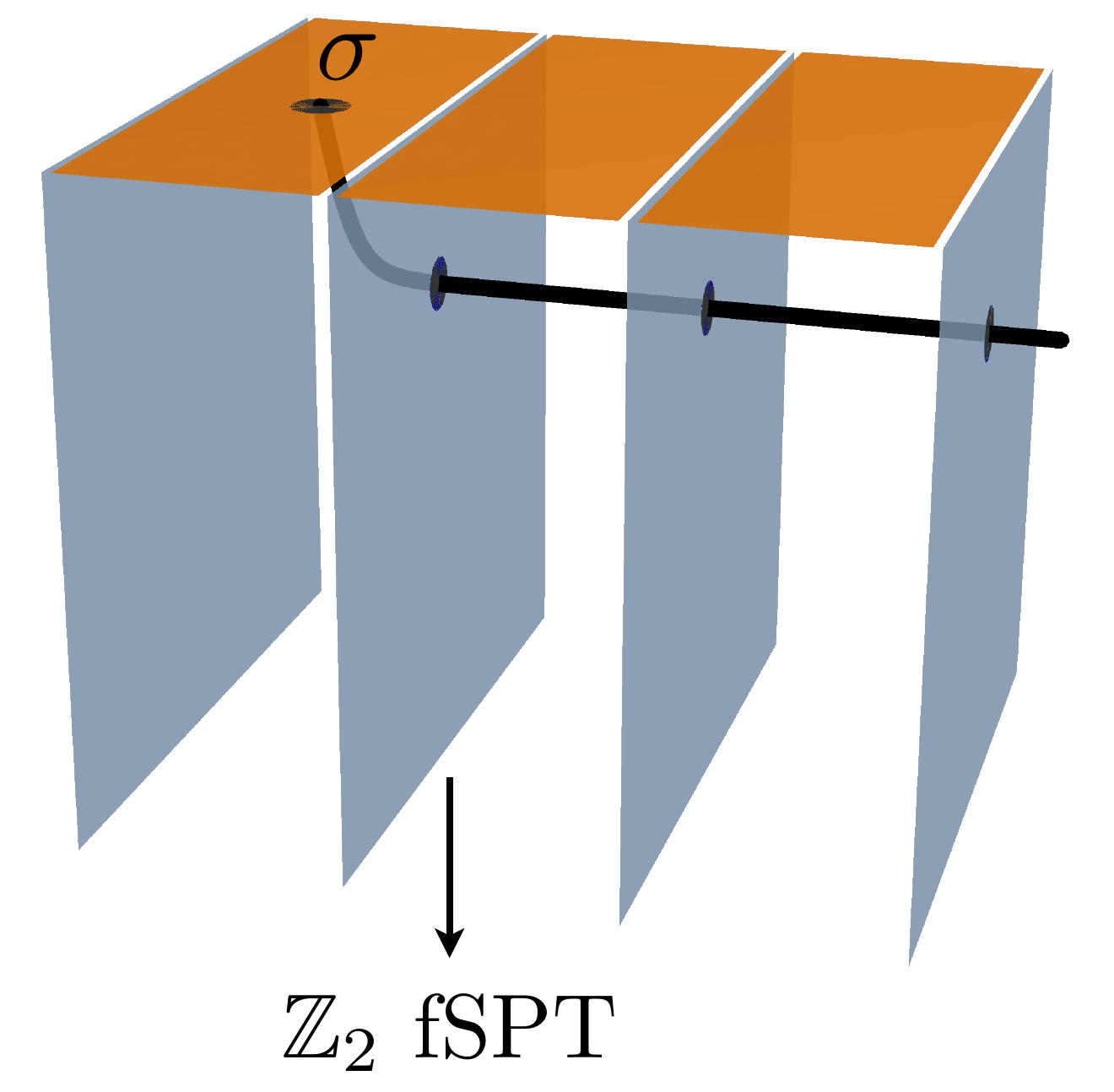}
	\caption{Illustration of the 3D weak fermionic $\Z_2$ SPT phase. A fermion parity flux line terminates on the surface. Each intersection of the flux line with the 2D SPT layers carries a half $\Z_2$ charge.}
	\label{fig:stack}
\end{figure}

These inconsistencies can be ``resolved'' by taking into account the weak SPT bulk. In the case of gauging fermion parity, the actual fermion parity flux has to extend into the bulk, so the flux line penetrates (an infinite number of) layers of 2D $\Z_2$ SPT phases. See Fig. \ref{fig:stack} for an illustration. It is known that for each layer, a fermion parity flux carries a half $\Z_2$ charge~\cite{Cheng_fSPT}. If we translate the fermion parity flux by one unit spacing, we change the number of layers penetrated by the flux line by one, which exactly compensates the anomalous $T_y$ action on the surface. 

With this bulk perspective, we can also resolve the anomaly in the fusion rules of $\Z_2$ symmetry defects $0_\mb{g}$ discussed earlier. Again the defect line extends indefinitely into the bulk. When we consider the two ways of computing $\rho_{T_y}(0_\mb{g}^2)$, they actually differ by a bulk process, namely fusing two (identical) $\Z_2$ defects on one of the 2D SPT layers. See Fig. \ref{fig:weak1} for an illustration. This process contributes another $f$ to compensate the mismatch on the surface, because in the 2D fermionic SPT phase we have $0_\mb{g}^2=f$~\cite{SET, Cheng_fSPT}. An intuitive way to see why this is the case is to momentarily enlarge the $\Z_2$ symmetry to $\U$, and $0_\mb{g}$ is just the $\pi$ flux in the Chern insulator where $\Z_2$ acts nontrivially. Then $0_\mb{g}^2$ amounts to a $2\pi$ flux insertion, which nucleates a fermion due to the quantum Hall effect.

We are now ready to derive a general constraint on the surface SET with this bulk. Motivated by the example, let us consider generally ${}^{T_y}(a_\mb{g}\times b_\mb{g})$. The bulk-surface relation requires that
\begin{equation}
	{}^{T_y}(a_\mb{g}\times b_\mb{g})= {}^{T_y}a_\mb{g}\times {}^{T_y}b_\mb{g} \times f.
	\label{}
\end{equation}
Here ${}^{T_y}x_\mb{g}$ denotes the $T_y$ transformations of defects in a SET that differs from the ``reference'' one by a fractionalization class $[\coho{w}]$. For clarity, we assume that the reference one is non-anomalous, and can be realized in purely 2D systems. First, we have
\begin{equation}
	\begin{split}
		{}^{T_y}a_\mb{g}\times {}^{T_y}b_\mb{g} &=\coho{b}(T_y, \mb{g})\times\rho_{T_y}(a_\mb{g})\times \coho{b}(T_y, \mb{g})\times \rho_{T_y}(b_\mb{g})\\
		&=\coho{b}(T_y, \mb{g})\times {}^\mb{g}\coho{b}(T_y, \mb{g})\times\rho_{T_y}(a_\mb{g})\times  \rho_{T_y}(b_\mb{g})\\
		&=\coho{b}(T_y, \mb{g})\times {}^\mb{g}\coho{b}(T_y, \mb{g})\\
		&\quad\times\coho{w}(\mb{g,g}) \times\sum_{c_\mb{0}}N_{\rho_{T_y}(a_\mb{g}),\rho_{T_y}(b_\mb{g})}^{c_\mb{0}}c_\mb{0}\\
		&=\coho{b}(T_y, \mb{g})\times {}^\mb{g}\coho{b}(T_y, \mb{g})\times\coho{w}(\mb{g,g})
		\times\sum_{c_\mb{0}}N_{a_\mb{g},b_\mb{g}}^{ {}^{T_y^{-1}}c_\mb{0}}c_\mb{0}\\
		&=\coho{b}(T_y, \mb{g})\times {}^\mb{g}\coho{b}(T_y, \mb{g})\times\coho{w}(\mb{g,g})\times\sum_{c_\mb{0}}N_{a_\mb{g},b_\mb{g}}^{ c_\mb{0}} {}^{T_y}c_\mb{0}.
	\end{split}
	\label{eqn:xyz}
\end{equation}
Notice that when going from the second to the third lines, we deliberately use
\begin{equation}
	\rho_{T_y}(a_\mb{g})\times \coho{b}(T_y, \mb{g}) = {}^\mb{g}\coho{b}(T_y, \mb{g})\times\rho_{T_y}(a_\mb{g}),
	\label{}
\end{equation}
which is also equal to $\coho{b}(T_y, \mb{g})\times\rho_{T_y}(a_\mb{g})$. At this point it is ambiguous. The reason we use ${}^\mb{g}\coho{b}(T_y, \mb{g})\times\rho_{T_y}(a_\mb{g})$ will be justified a bit later. In the manipulation, a crucial step is to use the symmetry property of defect fusion rules:
\begin{equation}
	N_{\rho_{T_y}(a_\mb{g}), \rho_{T_y}(a_\mb{h})}^{c_\mb{gh}} = N_{a_\mb{g}, b_\mb{h}}^{\rho_{T_y}^{-1}(c_\mb{gh})}.
	\label{}
\end{equation}
It follows from our choice of a non-anomalous $\rho$ as the reference point.

On the other hand,
\begin{equation}
	\begin{split}
		 {}^{T_y}(a_\mb{g}\times b_\mb{g})&=
		 {}^{T_y}[\coho{w}(\mb{g,g})\times\sum_{c_\mb{0}}N_{a_\mb{g}b_\mb{g}}^{c_\mb{0}}c_\mb{0}]	\\
		 &={}^{T_y}\coho{w}(\mb{g,g})\times\sum_{c_\mb{0}}N_{a_\mb{g}b_\mb{g}}^{c_\mb{0}} {}^{T_y}c_\mb{0}.
	\end{split}
	\label{}
\end{equation}
Comparing the two calculations, the bulk-surface correspondence requires
\begin{equation}
	{}^{T_y}\coho{w}(\mb{g,g}) = \coho{b}(T_y, \mb{g})\times {}^\mb{g}\coho{b}(T_y, \mb{g})\times\coho{w}(\mb{g,g}) \times f.
	\label{eqn:type-I}
\end{equation}
One can check that both sides are invariant under coboundary transformations. Had we used $\coho{b}(T_y, \mb{g})\times\rho_{T_y}(a_\mb{g})$ instead of ${}^\mb{g}\coho{b}(T_y, \mb{g})\times\rho_{T_y}(a_\mb{g})$ in Eq. \eqref{eqn:xyz}, this would no longer be true.

Similar to the fermionic LSM-type constraints, here Eq. \eqref{eqn:type-I} also demands that $T_y$ must permute anyons. Otherwise we would have $\coho{b}(T_y, \mb{g})\times {}^\mb{g}\coho{b}(T_y, \mb{g})=f$, or $\overline{\coho{b}(T_y, \mb{g})}={}^\mb{g}\coho{b}(T_y, \mb{g})\times f$, but the two sides have opposite topological twist factors: $\theta_{\overline{\cohosub{b}}}=\theta_{\cohosub{b}}$, but the right-hand side $\theta_{\cohosub{b}\times f}=-\theta_{\cohosub{b}}$.

\begin{widetext}
	\begin{center}
\begin{figure}[ht]
	\centering
	\includegraphics[width=0.85\textwidth]{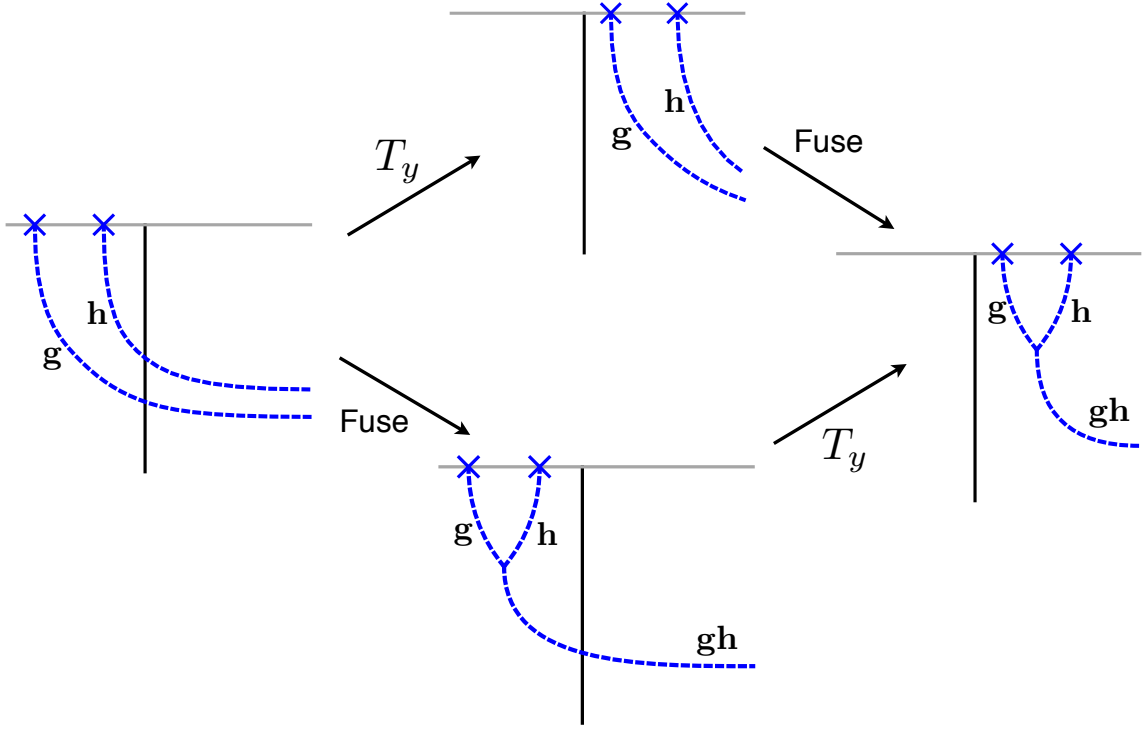}
	\caption{Illustration of the defect fusion anomaly, as a side view of Fig. \ref{fig:stack}. The vertical solid line represents a 2D fermionic SPT layer, and dashed lines are defect lines, terminating on the surface (the horizontal line).  One may first translate the two surface defects, and then fuse them, or first fuse them, then translate. The difference between the two paths is given by the projective defect fusion rules on the 2D fermionic SPT layer.}
	\label{fig:weak1}
\end{figure}
\end{center}

\end{widetext}

It is straightforward to generalize this result to generic type-I weak fermionic SPT phases. Suppose the internal symmetry group is $G$. The 2D fermionic SPT phase per unit length falls within the group super-cohomology classification~\cite{GuWen}, characterized by a $2$-cocycle $[\nu]\in \H^2[G, \Z_2]$. More physically, the $G$ defects have projective fusion rules determined by $\nu$.    The bulk-boundary correspondence then implies
\begin{equation}
	{}^{T_y}a_\mb{g}\times {}^{T_y}b_\mb{h} = \nu(\mb{g,h})\times {}^{T_y}(a_\mb{g}\times b_\mb{h}).
	\label{}
\end{equation}

We now evaluate the two sides of the equation. First we have
\begin{equation}
	\begin{split}
		{}^{T_y}a_\mb{g}\times {}^{T_y}b_\mb{h} &=\coho{b}(T_y, \mb{g})\times\rho_{T_y}(a_\mb{g})\times \coho{b}(T_y, \mb{h})\times \rho_{T_y}(b_\mb{h})\\
		&=\coho{b}(T_y, \mb{g})\times {}^\mb{g}\coho{b}(T_y, \mb{h})\times\rho_{T_y}(a_\mb{g})\times  \rho_{T_y}(b_\mb{h})\\
		&=\coho{b}(T_y, \mb{g})\times {}^\mb{g}\coho{b}(T_y, \mb{h})\times\coho{w}(\mb{g,h})\\
		&\quad\times\sum_{c_\mb{gh}}N_{\rho_{T_y}(a_\mb{g}),\rho_{T_y}(b_\mb{h})}^{c_\mb{gh}}c_\mb{gh}\\
		&=\coho{b}(T_y, \mb{g})\times {}^\mb{g}\coho{b}(T_y, \mb{h})\times\coho{w}(\mb{g,h})\\
		&\quad\times\sum_{c_\mb{gh}}N_{a_\mb{g},b_\mb{h}}^{ c_\mb{gh}} \rho_{T_y}(c_\mb{gh}).
	\end{split}
	\label{}
\end{equation}
On the other hand,
\begin{equation}
	\begin{split}
		 {}^{T_y}(a_\mb{g}\times b_\mb{h})&=
		 {}^{T_y}\big[\coho{w}(\mb{g,h})\times\sum_{c_\mb{gh}}N_{a_\mb{g},b_\mb{h}}^{c_\mb{gh}}c_\mb{gh}\big]	\\
		 &={}^{T_y}\coho{w}(\mb{g,h})\times\sum_{c_\mb{gh}}N_{a_\mb{g},b_\mb{h}}^{c_\mb{gh}} {}^{T_y}c_\mb{gh}\\
		 &={}^{T_y}\coho{w}(\mb{g,h})\times \coho{b}(T_y, \mb{gh})\times\sum_{c_\mb{gh}}N_{a_\mb{g},b_\mb{h}}^{c_\mb{gh}} \rho_{T_y}(c_\mb{gh}).
	\end{split}
	\label{}
\end{equation}
Comparing the two calculations, we finally obtain the following condition on the fractionalization class:
\begin{multline}
	{}^{T_y}\coho{w}(\mb{g,h}) \times \coho{b}(T_y, \mb{gh}) \\
	= \coho{b}(T_y, \mb{g})\times {}^\mb{g}\coho{b}(T_y, \mb{h})\times \coho{w}(\mb{g,h}) \times \nu(\mb{g,h}).
	\label{}
\end{multline}

\section{Discussions}
\label{sec:discussion}
In \Ref{ChengPRX2016} it was shown that the classification of weak bosonic SPT phases, as well as the corresponding LSM-type constraints in 2D systems can be obtained by formally treating the translation symmetry as an internal $\mathbb{Z}$ symmetry. More specifically, the LSM constraint in (bosonic) SET phases was derived by matching the $\H^4$ obstruction class~\cite{Chen2014} of the 2D theory, with the $\H^4$ cocycle which characterizes the bulk group-cohomology SPT phase. In \Ref{ThorngrenPRX2018} a more general ``crystalline equivalence principle'' was formulated, relating topological classifications of phases of matter with crystalline symmetries and those with internal symmetries having the same abstract group structure.

One may wonder whether the results obtained in this work can also be applied to fermionic SPT phases with internal symmetries. A partial classification of such fermionic SPT phases in 3D was proposed in \Ref{GuWen}, known as the group-superchomology construction, where the data is a $\Z_2$-valued $3$-cocycle $[\rho]\in \H^3[G, \Z_2]$. 
The fermionic LSM theorem for $\Z_2$ can be viewed as a surface constraint of a 3D fermionic SPT phase with $G=\Z\times \Z\times \Z_2$ symmetry. For an explicit expression of the cocycle, denote the group elements of $\Z\times \Z\times \Z_2$ additively as $\mb{a}=(a_1,a_2,a_3)$ where $a_1,a_2\in \Z$ and $a_3=0,1$, then
\begin{equation}
	\rho_{\text{II}}(\mb{a},\mb{b},\mb{c})= a_1b_2c_3\text{ mod }2.
	\label{}
\end{equation}
Similarly, for type-I weak SPT phases, denote elements of $\Z\times \Z_2$ by $(a_1, a_2)$,
\begin{equation}
	\rho_\text{I}(\mb{a,b,c})=\frac{1}{2}a_1(b_2+c_2-[b_2+c_2]_2)\text{ mod }2.
	\label{}
\end{equation}
In \Ref{Cheng3DFSPT} it was shown that the $\Z$ group can be replaced with a simpler $\Z_4$ group in the presence of strong interactions. Namely, there exists an interacting fermionic SPT phases in 3D protected by $\Z_4\times \Z_2$ symmetry. In the type-I weak SPT picture, $\Z_4$ can be understood intuitively as follows: if we were to gap out the surface by spontaneously breaking the translation symmetry, the minimal enlargement of unit cells is $4$ (thus $\Z\rightarrow \Z_4$) since the $\Z_2$ 2D fermionic SPT per unit length is $\Z_4$-classified. 
It will be interesting to have a more systematic understanding of the general constraints on surface topological order for group super-cohomology SPT phases.

In this work, we assume that the fermions transform linearly under the internal symmetry group, i.e. the symmetry group takes the form $\Z_2^f\times G$. More generally fermions may transform projectively. It is an interesting question to see how our results generalize. In addition, we have not addressed the LSM-type theorems with time-reversal symmetries. In this case, our approach, which relies crucially on symmetry defects, does not apply. New methods will have to be developed to handle anti-unitary symmetries. One potential solution is to again exploit the crystalline equivalence principle and instead consider spatial reflection symmetries.

Recently, \Ref{WangGu2017} and \Ref{ThorngrenKapustin2017} proposed a new class of interacting fermionic SPT phases with $\Z_2^f\times G$ symmetry in 3D, beyond the super-cohomology classification. They are classified by a $2$-cocycle $[\sigma]\in\H^2[G, \Z_2]$ (subject to certain obstruction-vanishing conditions). The physics of these SPT phases is that Majorana chains are decorated on ``line junctions'' of domain walls (when $\mb{g}, \mb{h}$ domain walls fuse into a $\mb{gh}$ domain wall). We believe that the Majorana LSM-type theorem mentioned in Sec. \ref{sec:majorana} can be viewed as the surface of such SPT with $G=\Z\times \Z$ (indeed $\H^2[\Z\times\Z, \Z_2]=\Z_2$, which is basically saying there is a Majorana mode per unit cell). Finding constraints on surface SET phases in this case is an interesting question and will be left for future work. 

Overall, an important task is to develop a general theory of fermionic SET phases. We have made partial progress along this direction in this work, but certain important aspects are left out. For example, when classifying topological symmetries we only consider $\text{Aut}(\mathcal{C}_0)$, but in fact the whole $\text{Aut}(\mathcal{C})$ of the modular extension needs to be considered. This is closely related to the surface anomaly of 3D fermionic SPT phases with a nontrivial $[\sigma]$.

We have mainly focused on the theoretical implications of fermionic LSM-type theorems in symmetric gapped phases. While we provide an example of such phase using coupled wire construction, the model is clearly unrealistic and designed to enable analytical solutions. In the future an important direction is to numerically study more realistic models of interacting spinless fermions, in particular with suitable interactions to frustrate symmetry-breaking orders.

\section{Acknowledgement}
I am grateful to Chao-Ming Jian, Chenjie Wang and Dominic Williamson for invaluable discussions and collaborations on related topics. I would like to thank Chinese University of Hong Kong for hospitality while part of the work was performed. This work is supported by startup funds from the Yale University.

\emph{Note added:} While the manuscript is being finalized, \Ref{Fidkowski2018} appeared on arXiv which proposed a general 't Hooft anomaly for surface states of group super-cohomology SPT phases. Our results are consistent with the more general anomaly matching conditions in \Ref{Fidkowski2018}.

\appendix

\section{Structure of Abelian anyons}
\label{app:abelian}
We will prove that an arbitrary Abelian braided fusion category with a transparent fermion $f$ can be written as $\mathcal{A}=\Z_2^f\times \tilde{\mathcal{A} }$. Here, a transparent fermion $f$ means $\theta_f=-1, f^2=1$ and for any $a\in\mathcal{A}$ we have $M_{a,f}=1$. Denote $\Z_2^f=\{1,f\}$.

Generally, $\Z_2^f$ is a normal subgroup of $\mathcal{A}$, so we can always view $\mathcal{A}$ as an extension of $\tilde{\mathcal{A}}=\mathcal{A}/\Z_2^f$. Such extensions are classified by $\H^2[\tilde{\mathcal{A}}, \Z_2^f]$. Suppose that $\tilde{\mathcal{A}}=\prod_{i=1}^K \Z_{N_i}$. Then
\begin{equation}
	\H^2[\tilde{\mathcal{A}}, \Z_2^f]=\prod_{i=1}^K \Z_{(N_i,2)} \prod_{1\leq i<j\leq K} \Z_{(N_i, N_j,2)}.
	\label{}
\end{equation}
First, observe that if $N_i$ is odd, it does not contribute to the second cohomology.  So we can assume that all $N_i$ are even.  Let us write down explicit cocycles. Denote an element of $\tilde{A}$ by a tuple $(x_1,x_2, \cdots, x_K)$ where $x_i$ are integers mod $N_i$, and group multiplication is denoted additively. A general cocycle can be written as
\begin{equation}
	\omega(x, y) = \sum_i \frac{p_{i}}{N_i}(x_i+y_i-[x_i+y_i]_N) + \sum_{ij}q_{ij}x_i y_j.
	\label{}
\end{equation}
Here $p_i, q_{ij}$ are integers. It is understood that $\omega(x,y)$ is defined mod $2$. If $q_{ij}\neq 0\, (\text{mod }2)$, then the corresponding extended group is non-Abelian. So we can set $q_{ij}=0$. The remaining part of the cocycle decouples for each $i$. Let us focus on one of the $\Z_{N_i}$ subgroup.

  Denote the generator of $\Z_{N_i}$ by $a_i$. The nontrivial cohomology class $p_i=1$ corresponds to 
\begin{equation}
	a_i^{N_i}=f.
	\label{}
\end{equation}
Then we must have $\theta_{a_i}^{N_i^2}=-1$. We can then compute the braiding statistics between $a_i$ and $f$ using the ribbon identity:
\begin{equation}
	M_{a_i^{N_i/2},f}=\frac{\theta_{a_i^{N_i/2}\times f}}{\theta_{a_i^{N_i/2}}\theta_f}=-\frac{\theta_{a_i^{3N_i/2}}}{\theta_{a_i^{N_i/2}}}=-\theta_{a_i}^{2N_i^2}=-1,
	\label{}
\end{equation}
contradicting the fact that $f$ is transparent. Therefore, the braiding structure fixes the cohomology class to be the trivial one, thus $\mathcal{A}=\tilde{\mathcal{A}}\times \Z_2^f$.

In case that $\mathcal{C}=\mathcal{A}$, i.e. $\mathcal{A}$ is super-modular, it immediately follows that $\tilde{\mathcal{A}}$ is a modular tensor category.

\bibliography{refs}

\end{document}